\documentclass[12pt]{article}

\usepackage[utf8]{inputenc}
\usepackage{enumitem}
\usepackage[a4paper,margin=1in]{geometry}

\usepackage{amssymb}
\usepackage{amsmath}

\usepackage{graphicx}
\usepackage{caption}
\usepackage{subcaption}
\usepackage{float}
\usepackage{multirow}
\usepackage{makecell}
\usepackage{tabularx}
\usepackage{booktabs}

\usepackage{hyperref}
\usepackage[authoryear]{natbib}

\title{How did the Urban Network Flow Adapt to the Collapse of the Carola Bridge?}

\author{
Jyotirmaya Ijaradar$^{1}$ \and
Ning Xie$^{1}$ \and
Lei Wei$^{1}$ \and
Sebastian Pape$^{1}$ \and
Matthias Körner$^{1}$ \and
Meng Wang$^{1}$ \and
\\
$^{1}$Chair of Traffic Process Automation, Technische Universität Dresden,\\
Dresden, Germany
}

\date{}

\begin{document}

\maketitle

\begin{abstract} The unexpected collapse of the Carola Bridge in Dresden, Germany, provides a rare opportunity to characterise how urban network traffic adapts to an unexpected infrastructure disruption. This study develops a data-driven analytical framework using traffic data from the Dresden traffic management system to assess the short-term impacts of the disruption. By combining statistical comparisons of pre- and post-collapse  motorised traffic distributions, peak-hour shifts, and Park-and-Ride data analyses, the framework reveals how traffic dynamics and traveller choices adjust under infrastructure disruption.  Results reveal that the two closest bridges, the Albert and Marien Bridges, absorb the majority of the diverted motorised traffic. In particular, the daily traffic volume on the Albert bridge increases by up to 81\%, which is equivalent to 3.5 hours of traffic operating with maximum flow.  Peak hours on critical links are significantly prolonged, reaching up to 250 minutes. Besides redistribution, the overall daily motorised traffic crossing the Elbe river declines by approximately 8,000 vehicles, while Park-and-Ride usage increases by up to 188\%, suggesting a potential travel mode shift after the disruption. The study reveals the patterns of traffic redistribution following an unexpected disruption and provides insights for resilience planning and emergency traffic management.
\end{abstract}

\noindent \textbf{Keywords:} Road network disruption; Critical infrastructure; Route choice; Travel mode choice; Traffic resilience


\section{Introduction}
\label{sec1}


On September 11, 2024, at approximately 2:59 AM, the Carola Bridge over the Elbe River in Dresden collapsed. Although the incident resulted in no injuries, it disrupted a vital multimodal corridor connecting Dresden’s Altstadt (Old Town) and Neustadt (New Town). The Carola Bridge had been one of the four main Elbe crossings in downtown Dresden, and a lifeline for the city’s transport network. The sudden collapse disrupted roughly 30,000–32,000 daily motorised traffic, two tram lines, and affected cyclists and pedestrians who used the bridge. This unexpected infrastructure failure provides a rare real-world opportunity to analyse how an urban transport network adapts to a sudden and severe disruption.



Disruptions in transport systems can be broadly classified as either expected or unexpected \citep{Mueller2020,Zhu2011}. Expected disruptions, such as roadworks \citep{Jamous2018}, public transport strikes \citep{Exel2001, Yang2022_transgeo}, or major public events \citep{AnsariEsfeh2022}, are known in advance. This allows travellers to adapt their plans and respond in a more structured manner, even though prolonged travel time is unavoidable due to increased demand-capacity mismatch. 
In contrast, unexpected disruptions, such as bridge collapses \citep{zhu2010traffic}, tunnel failures \citep{Tympakianaki2018}, natural disasters \citep{Rahman2025, Yonson2020_trans_geo, ALAM2024103852_transgeo}, or major accidents \citep{SanchezGonzalez2021}, occur without warning, forcing travellers to make immediate changes to their travel plans. These events lead to uneven traffic redistribution and increase congestion on alternative routes, resulting in prolonged disequilibrium across the network \citep{danczyk2017unexpected}. A prominent example is the I-35W bridge collapse in Minneapolis in 2007. The incident interrupted over 140,000 daily vehicle trips and led to significant shifts in traffic flows and commuter behaviour in the  weeks after the collapse \citep{zhu2010traffic, he2012modeling}. Similarly, the fire-induced collapse of the I-95 overpass in Philadelphia caused severe traffic disruptions and highlighted the urgent need to incorporate fire resilience in bridge design to prevent sudden failures of critical transport infrastructure \citep{Kodur2024}. More recently, the collapse of the Francis Scott Key Bridge caused weeks-long traffic disruptions at the Port of Baltimore, severely impacting global supply chains \citep{Gracia2024}. These examples illustrate that unexpected bridge failures can disrupt not only daily commutes but also commerce and regional mobility, with impacts lasting from days to months.

Empirical studies on network disruption focused on infrastructure resilience \citep{Liu2023}, network vulnerability \citep{Fikar2016_transgeo}, or traffic rerouting strategies \citep{Siri2022, Ezaki2022}. For instance, \cite{Chang2001} studied the impacts of the Loma Prieta, Northridge, and Kobe earthquakes on transport systems by measuring how much of the network remained open and accessible. In addition, \cite{Sumalee2006} focused on creating optimal traffic control plans to be used after a major disaster. Similar studies have been performed in recent years in the 2021 British Columbia floods \citep{Liao2024} and the Ahr valley flood \citep{Wassmer2024}. These models often assume idealized user behaviour (e.g., user equilibrium) and neglect the stochastic, day-to-day behavioural adaptation. 

Prior studies usually emphasized long-term recovery and traveller behaviour model, the empirical analysis of short-term traffic pattern evolution within the first few weeks remains less understood. This gap is largely due to the limited availability of real-time data during such disruptive events \citep{Nalin2025}. Although the study of the collapse of the I-35W bridge  is the most detailed empirical work on this topic to date, the main focus is on motorway traffic \citep{zhu2010traffic}.
Although the study of \cite{Szarata2019} provides some insights into the traffic disruption in a European city, its reliance on data from only selected bridge crossings and subjective resident surveys may not fully capture the complete, system-wide impact on the network. The collapse of the Carola Bridge, therefore, presents a unique opportunity to study how travellers respond to a sudden network failure in a European urban context. 

This study aims to address the knowledge gap regarding short-term traffic disruptions from the perspective of a European city. Unlike previous studies, which either focused on structural failures or relied on limited survey or single corridor data, our analysis leverages network traffic data to capture the system-wide redistribution of flows and multimodal adaptations across the entire network. To guide our investigation, we formulate the following research questions:

\begin{itemize}[itemsep=0pt, topsep=2pt]
       \item 1) What are the spatial-temporal effects on traffic flow redistribution across the network after the bridge collapse?
     \item 2) How do peak-hour timing and intensity change in daily traffic pattern after the network disruption?
    \item 3) How do travellers adjust their travel behaviour and mode choice in response to the network disruption?
\end{itemize}


To address these research questions, we adopt a data-driven analytical framework based on traffic flow data collected from the VAMOS Traffic Management System in Dresden, Germany \citep{DresdenRoadsOffice}. For the network-level analysis, post-collapse traffic data are first compared with a pre-collapse baseline using paired T-tests and ANOVA tests. Then, an Equivalent Full-Capacity Hours (EFCH) metric is developed to quantify changes in traffic flow dynamics and capacity utilisation across the network, thereby capturing the spatio-temporal redistribution of traffic flow. After that, a rolling Peak Hour Indicator (PHI) is developed to identify shifts in the timing and intensity of daily traffic peaks, allowing variations in daily traffic patterns to be examined. Besides, commuter behavioural responses are assessed using Park-and-Ride (P+R) data combined with motorway node data, which  identify travel mode shifts and route diversions that reflect adaptive travel behaviour after the disruption. Through these analyses, the proposed  research questions are addressed.

The remainder of the paper is organized as follows: Section~\ref{sec:data-methods} describes the dataset and methodology, including the EFCH metric, PHI and statistical assessment techniques. Section~\ref{sec:results} presents empirical results from the pre- and post-collapse periods and discusses behavioural changes. Finally, Section~\ref{sec:conclusion} concludes the paper and outlines directions for future research.

\section{Data-driven analytical framework }\label{sec:data-methods}


The empirical study was conducted using a structured, data-driven framework designed to examine changes in traffic conditions following the Carola Bridge collapse. The methodology integrates multiple analytical components, including overall network-level assessment of traffic flow redistribution, daily peak hour analysis, and commuter behavioural response evaluation. For this purpose, we identified the main Elbe River crossings, constructed robust pre- and post-collapse baselines, and performed a systematic data-driven analysis using traffic flow data from the VAMOS Traffic Management System. The following subsections describe in detail the analytical framework and the systematic data processing procedures applied in this study.




\subsection{Data and Processing}
\begin{figure}
    \centering
    \includegraphics[width=1\linewidth]{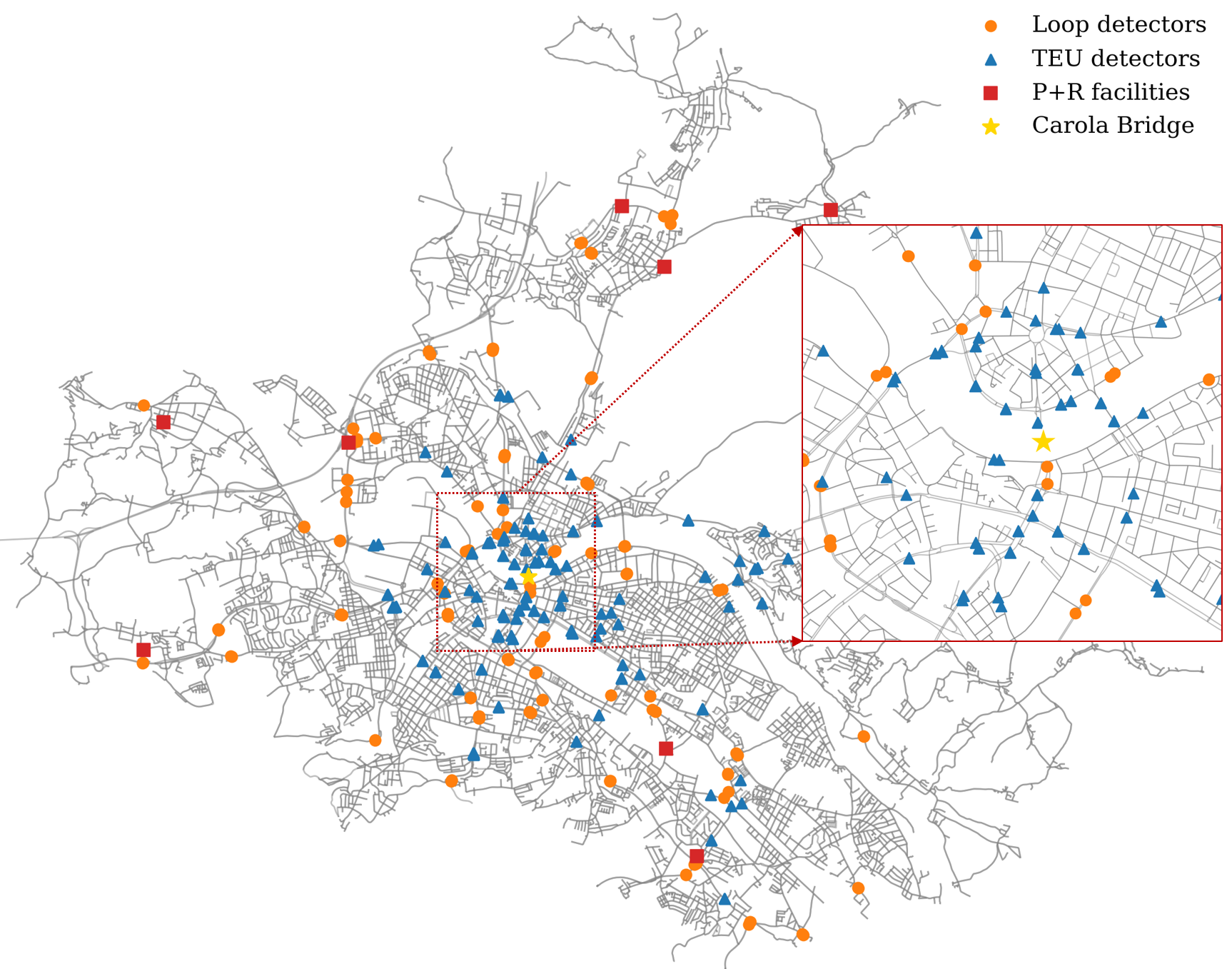}
    \caption{Overview of the VAMOS sensor network in Dresden, showing the locations of both traffic flow and parking occupancy detectors. The location of the Carola Bridge is highlighted, and the inset provides a detailed view of the inner-city area surrounding the bridge.}
    \label{fig:vamos_sensor_overview}
\end{figure}

The primary data source for this study is the VAMOS Traffic Management System in Dresden, which provides real-time traffic measurements through a network of over 1,800 sensors comprising inductive loops, Traffic Eye Universe (TEU) cameras, parking detectors, and roadside units (RSUs). Among these, 186 double induction loop detectors and 160 TEU detectors were selected for analysis \citep{Yan2024}. Near the Carola Bridge, VAMOS is equipped with double induction loops and infrared sensors, covering adjacent bridges and connecting roads. Figure \ref{fig:vamos_sensor_overview} shows the list of sensors located around the incident zone. These sensors collect data on traffic volume and average vehicle speeds at one-minute or five-minute intervals. For our analysis, we selected a comparative window of six weeks, where three weeks before and three weeks after the collapse of the Carola Bridge on September 11, 2024. This time frame provides sufficient temporal resolution to observe both immediate and stabilizing effects of the disruption.

In addition, we use parking occupancy data from designated P+R facilities. The VAMOS system is equipped with 46 parking detectors, which record occupancy data at an average interval of 15 minutes. Among these eight are located at P+R facilities: P+R Cossebaude, P+R Gompitz, P+R Grenzstraße, P+R Kaditz, P+R Klotzsche, P+R Langebrück, P+R Prohlis, and P+R Reick. Figure \ref{fig:vamos_sensor_overview} (red rectangles) illustrates the spatial distribution of all these P+R facilities.

\subsection{Identify Traffic Flow Redistribution}

For the starting of our empirical analysis, we focused on overall volume changes across the entire network and identified critical road segments for more detailed analysis. After the collapse, Dresden has seven alternative bridges for the river crossings: Elbe Bridge Motorway A4, Flügelweg Bridge, Marien Bridge, Augustus Bridge (Augustusbrücke), Albert Bridge, Waldschlößchen Bridge, and Loschwitzer Bridge. Among these, Augustus Bridge is restricted to trams and vulnerable road users (VRUs), while Elbe Bridge is reserved for motorway traffic and out of monitoring range; hence, both were excluded from the analysis. Additionally, no significant traffic volume changes were detected on the Loschwitzer Bridge during the observation period, so it was also excluded from further consideration. Based on the remaining river-crossing bridges, we constructed different traffic corridors and associated arteries for detailed analysis. 

To verify whether the changes are statistically meaningful, we used statistical methods to account for natural fluctuations in traffic patterns (e.g., daily, weekly, or seasonal changes). We applied the paired T-test \citep{Ross2017,Wang2022}  to compare traffic volumes before and after the disruption, as this method is commonly used to analyse the impact of events like COVID-19 \citep{Lannoy2022}, bridge collapses \citep{Zhu2011}, and natural disasters \citep{Avand2022}. The paired T-test evaluates whether the mean difference between two related samples differs significantly from zero. In addition, we used the ANOVA test \citep{Kim2017} to detect significant differences across critical road segments (e.g., bridges, corridors) and on different scenarios, which complements the paired T-test by providing a broader view of the changes observed \citep{Mishra2019, Wan2021}. For both tests, statistical significance is evaluated using a 5\% significance level. The mathematical formulations and test statistics for the paired T-test and ANOVA are provided in \ref{app:stat-analysis}.

To further evaluate the changes in daily traffic load or pressure experienced by individual road segments, we analyse the Equivalent Full-Capacity Hours (EFCH). This metric is based on the Maximum Observed Operational Flow (MOOF), defined as the 95th percentile of hourly volumes across all 15-minute intervals from both the pre- and post-collapse periods. MOOF values were computed separately for one-lane and two-lane roads and similar speed limits to reflect their differing flow capacities. For each category, the maximum MOOF observed in the network was used as a fixed benchmark. Using this reference, the EFCH is calculated by dividing the absolute change in daily traffic volume by the respective MOOF value, as shown below:

\begin{equation}
\text{EFCH}_i = \frac{\Delta Q_i}{\text{MOOF}_{R(i)}}
\end{equation}
\noindent
where $\Delta Q_i = Q^{\text{after}}_i - Q^{\text{before}}_i$ denotes the change in daily traffic volume at location $i$, with $Q^{\text{after}}_i$ representing the volume after the collapse and $Q^{\text{before}}_i$ representing the volume before the collapse. The $\text{MOOF}_{R(i)}$ is the MOOF value corresponding to the road category $R(i)$ (i.e., one-lane or two-lane). EFCH serves as a normalised indicator of traffic load across heterogeneous segments of the network. The results of the calculated MOOF are given in the appendix \ref{app:moof}.

\subsection{Peak-Hour Pattern Identification}

To fully understand the traffic dynamics throughout the day after the network disruption, daily traffic flow changes on critical roads are explored by comparing peak hour changes. Peak hour analysis is a common procedure, for example, using the Peak Hour Factor (PHF) \citep{HighwayManual2022}. However, the PHF from highway manuals is not very suitable for complex city networks, where some roads may already be at capacity before the disruption, and therefore relative changes in traffic flow may not reflect the real impact. To overcome this, we introduce a rolling index called the Peak Hour Indicator (PHI) to represent the time-varying traffic conditions for peak hour identification. Compared to traditional criteria, it combines normalised traffic flow and speed over a rolling window, as shown in Eq. (\ref{eq1}):
\begin{equation}
    PHI_t =\frac{1}{w}\sum_{i=0}^{w-1}\frac{\alpha_1\cdot\frac{q_{t-i}-q_{min,t}}{q_{max,t}-q_{min,t}}+\alpha_2\cdot(1-\frac{v_{t-i}-v_{min,t}}{v_{max}-v_{min,t}})}{\alpha_1+\alpha_2}
    \label{eq1}
\end{equation}
\noindent
where, $w$ is the rolling window size, $q_{t-i}$ is the traffic flow at time $t-i$, $q_{max,t}$ and $q_{min,t}$ are the maximum and minimum traffic flow during the rolling window at time $t$, $v_{t-i}$ is the traffic speed at time $t-i$, $v_{max,t}$ and $v_{min,t}$ are the maximum and minimum traffic speed during the rolling window at time $t$, and $\alpha_1$ and $\alpha_2$ are the weight coefficients of traffic flow and speed, respectively.

It is critical to determine the threshold for PHI based on the mean value, standard deviation, and $90^{th}$ percentile value over the whole day, which is formulated in Eq. (\ref{eq2}). The period during which the PHI is greater than the threshold is defined as peak hour. This formulation can be used to detect peak hours because PHI captures the simultaneous effect of high traffic flow and reduced speed over sustained periods, while the threshold ensures that only statistically significant deviations from normal daily conditions are classified as peak periods.
\begin{equation}
    \theta=\max(\frac{1}{N}\sum_tPHI_t+k\cdot\sqrt{\frac{1}{N}\sum_t(PHI_t-\frac{1}{N}\sum_tPHI_t)^2}, PHI_{90})
    \label{eq2}
\end{equation}
\noindent
where $\theta$ is the threshold of PHI for peak hour identification, $N$ is the total number of sampling times during the day, $PHI_{90}$ is the $90^{th}$ value of PHI, and $k$ is the weight coefficient of standard deviation.

\subsection{Behavioral Adaptation Indicators}

To analyse the commuter behaviour adaptation, we explore broader traveller responses in potential modal shifts and route choice. We can infer these behaviours from P+R data and motorway node data. Previous studies have shown that P+R data can be used for modal shift analysis \citep{Hamer2010, Zijlstra2015, MEHandayeni2018} and changes in motorway node data may reflect route choice changes \citep{Tiratanapakhom2013, Wang2019}. In Dresden, P+R parking is a special type of parking where it is free if motorists switch to public transport. Therefore, analysing the occupancy data can indicate whether travellers are shifting to public transport, which gives us an indirect insight into modal shifts. Additionally, we analysed data from motorway ramps and near-ramp areas to check for route choices and behavioural changes among regional commuters. In the following section, we discuss the different data used for this study.



\begin{figure}
    \centering
    \includegraphics[width=0.98\linewidth]{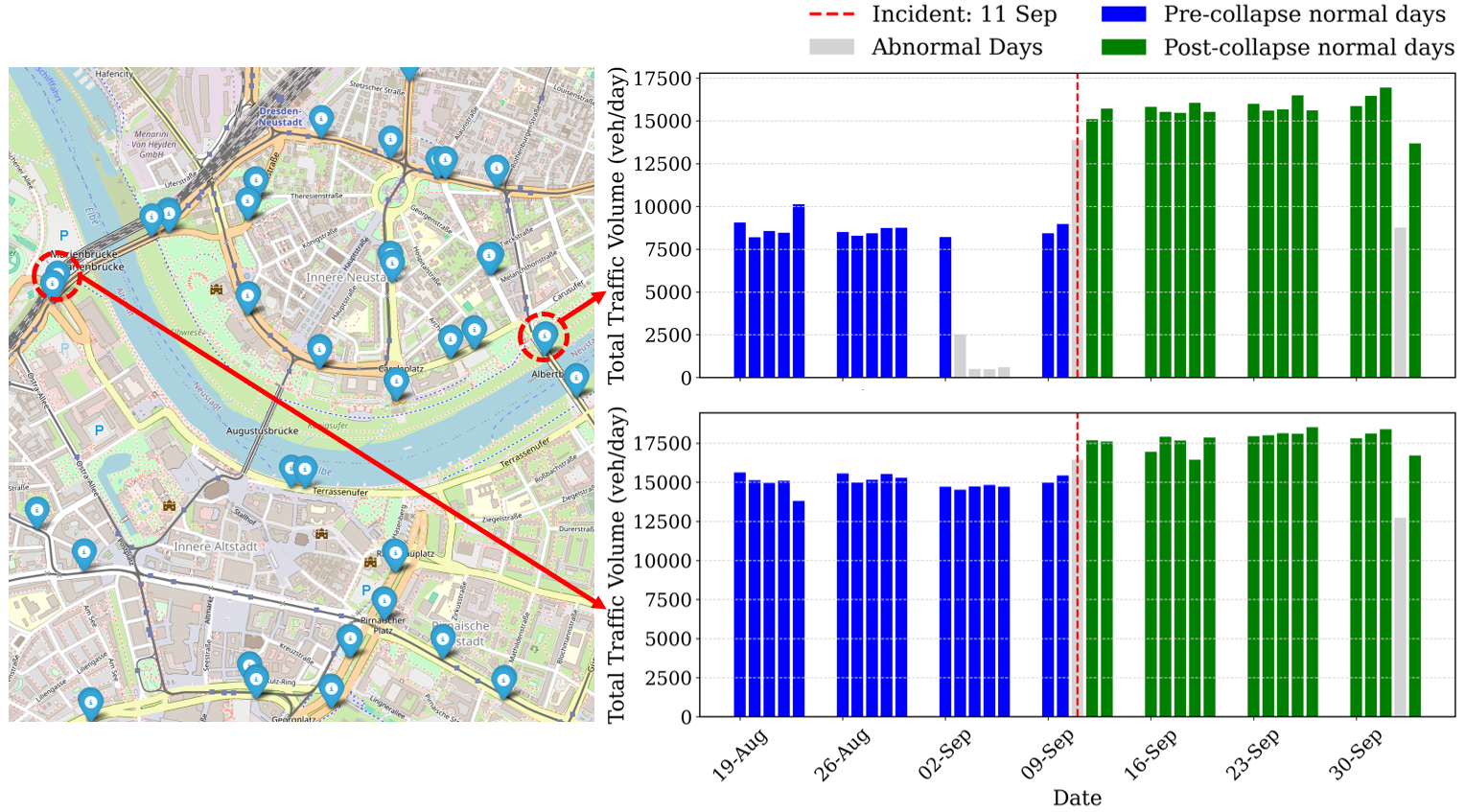}
    \caption{Daily traffic volume (veh/day) at two representative sensor locations (clockwise order with sensor IDs 1429\_1430 and 1407\_1408) around the Carola Bridge before and after the incident on 11 September. Grey bars indicate "abnormal" days, while blue and green bars represent “normal” reference days identified using DBSCAN clustering for pre- and post-collapse periods, respectively. The red dashed line denotes the day of the bridge collapse.}
    \label{fig:dbscan_result}
\end{figure}

To ensure the quality and consistency of the dataset, several preprocessing steps were undertaken. First, timestamps were standardised, and sensor readings with missing or invalid values were removed. Next, we employed the DBSCAN (Density-Based Spatial Clustering of Applications with Noise) algorithm \citep{Schubert2017} to identify and exclude anomalous days from the dataset. These outliers often correspond to public holidays, extreme weather events, or sensor malfunctions that could bias the analysis. By applying DBSCAN to daily traffic patterns (clustered by aggregated volume profiles), we isolated and removed days that did not conform to typical weekday behaviour. Following this filtering process, we selected “reference days” representing consistent, high-quality traffic conditions from both the pre- and post-collapse periods. These reference days serve as the empirical foundation for volume comparison, statistical testing, and further analysis such as peak hour analysis. Figure \ref{fig:dbscan_result} illustrates the DBSCAN clustering results for two representative sensor locations. In the figure, the blue and green bars indicate the selected error-free representative days for the pre- and post-collapse periods, respectively, while the grey bars represent all available data. The red dashed line marks the date of the incident on September 11th. The results clearly show that the DBSCAN algorithm effectively identifies and excludes anomalous or externally influenced days from the baseline data. For instance, in the top-right sensor data, the algorithm has omitted September 2nd, 3rd, 4th, and 5th from the pre-collapse baseline.

\section{Results and Discussion}\label{sec:results}

This section presents the empirical findings of the study and discusses the implications of the observed traffic changes after the Carola Bridge collapse using the proposed data-driven analytic framework. We begin by analysing the overall redistribution of traffic flows across the network and conducting statistical validation to quantify the spatialtemporal impacts of the disruption, thereby addressing research question 1. We then investigate changes in peak-hour timing and intensity, and analyse Park-and-Ride utilisation and motorway node flows to identify potential modal shifts and behavioural adaptations among travellers, thus addressing research questions 2 and 3.



\subsection{Analysis of Spatio-Temporal Effect}

\begin{figure}[!t]
    \centering
    \begin{subfigure}[t]{.75\textwidth}
        \centering
        \includegraphics[width=\textwidth]{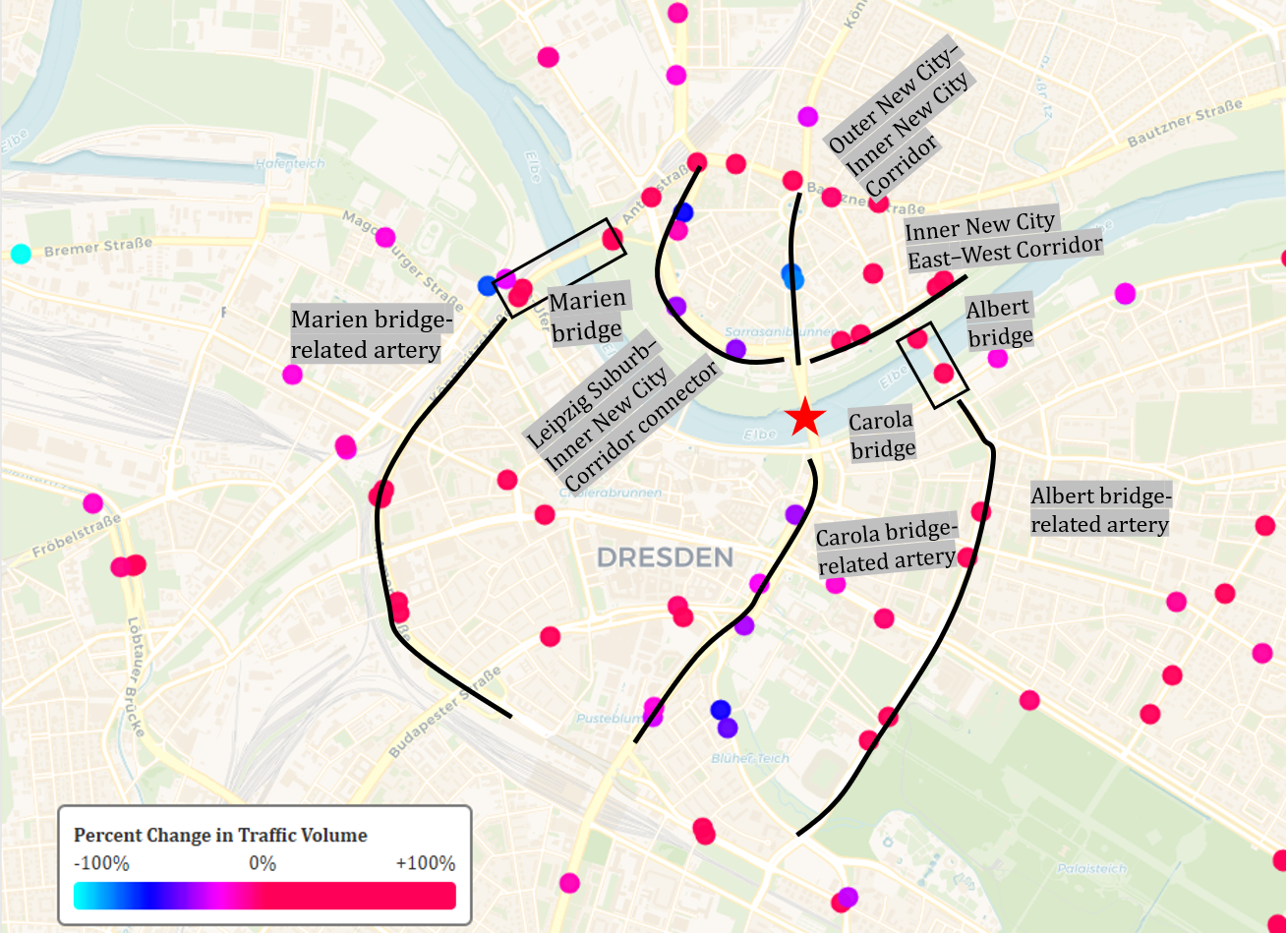}
        \caption{Percent change in traffic volume.}
        \label{phi before}
    \end{subfigure}
    \begin{subfigure}[t]{.75\textwidth}
        \centering
        \includegraphics[width=\textwidth]{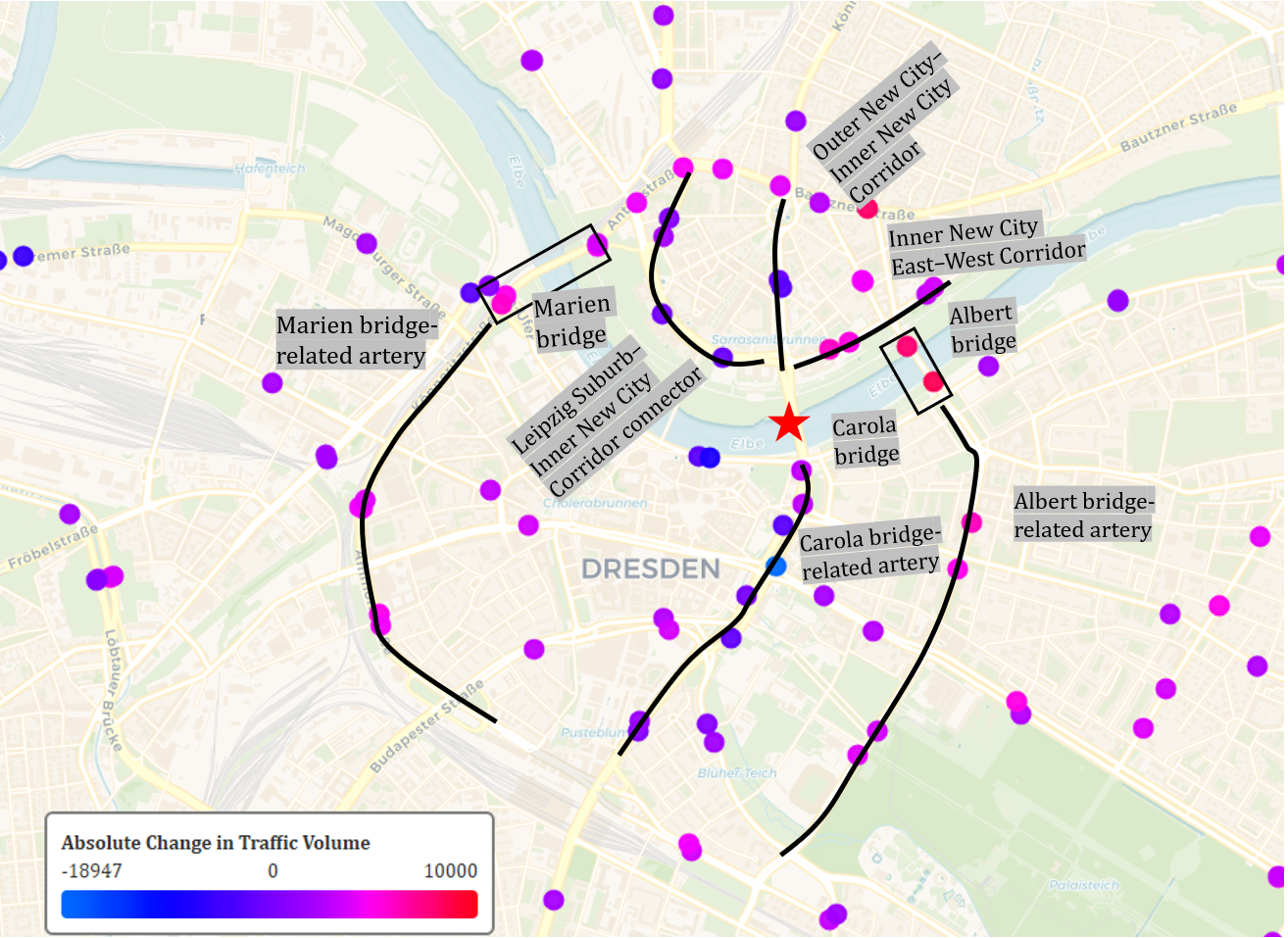
        }
        \caption{Absolute change in traffic volume.}
        \label{phi after}
    \end{subfigure}
    
    \caption{Daily traffic flow changes in Dresden after the Carola Bridge collapse. The alternative bridges are marked with rectangles, while the arteries and corridors are indicated with thick lines.}
    \label{fig:overall_chnages}
\end{figure}

\begin{figure}[!t]
    \centering
    \includegraphics[width=1\linewidth]{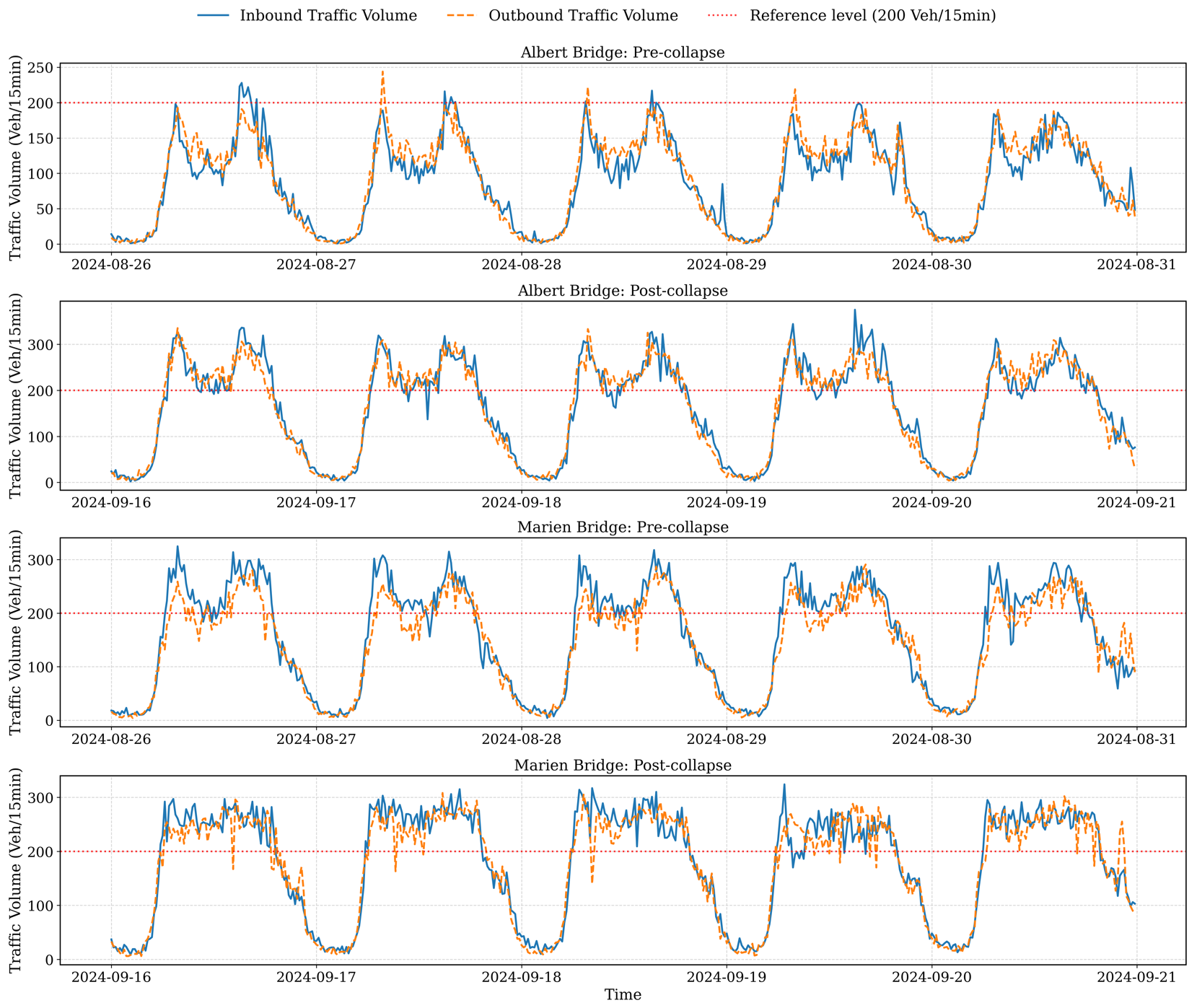}
    \caption{Time series of inbound and outbound traffic volumes (veh/15 min) at Albert Bridge and Marien Bridge before (week 35) and after (week 38) the Carola Bridge collapse.}
    \label{fig:flow-time-series}
\end{figure}

This section presents the detailed results of traffic volume changes following the collapse of the Carola Bridge. Figure \ref{fig:overall_chnages} visualises the overall spatial changes in daily traffic volume across Dresden in both relative and absolute terms. Our analysis reveals that the Albert Bridge and the Marien Bridge become the primary alternative river crossings, supplemented by the Flügelweg Bridge and Waldschlösschen Bridge as secondary routes. The figure \ref{fig:flow-time-series} illustrates changes in inbound and outbound traffic flows at Albert Bridge and the Marien Bridge before and after the Carola Bridge collapse for different weekdays. Post-collapse conditions show higher volumes and longer peak periods, demonstrating the role of primary alternative river crossings. Additionally, we select three key arterial roads and three distinct traffic corridors directly related to the Carola Bridge and the alternative bridges for a more detailed analysis.

Table~\ref{flow change} summarises the daily vehicle volumes before and after the collapse, including the absolute and relative changes for all detectors on these links. The table clearly shows that the Albert and Marien Bridges were the most impacted, both experiencing substantial increases in traffic as they absorbed diverted flows from the collapsed Carola Bridge. Specifically, Albert Bridge exhibits the most prominent increase, with traffic rising by approximately 6,849 and 7,048 vehicles per day in outbound and inbound directions, respectively, representing a high 77.49\% and 81.28\% surge. Marien Bridge also experiences notable growth, with outbound and inbound flows increasing by 3,345 and 2,738 vehicles per day, corresponding to a 24.20\% and 18.25\% rise, respectively. These results imply that the Albert Bridge probably replaces the role of the Carola Bridge.

The impacts extended beyond the bridges themselves, propagating through arterial roads and corridors. For instance, the Outer New City–Inner New City Corridor, which served as an outbound or inbound corridor to the Carola Bridge, recorded severe traffic reductions. Detector 1371 located on Albert street in the outbound direction towards Albertplatz shows a drop of 5,918 vehicles per day (-74.86\%), while the corresponding inbound detector 1322\_1323  towards Carolaplatz records a reduction of 5,390 vehicles per day (-71.08\%). Similarly, the Leipzig Suburb–Inner New City Corridor experiences traffic reductions ranging from 1,504 to 4,485 vehicles per day, with declines of up to -36.45\%, which is due to extended travel distance for river crossing.

In addition, arterial roads directly connected to the affected bridges also show a clear redistribution of traffic flow. Carola Bridge-related arteries experience drastic drops in flow, with detectors such as 1310\_1311 and 1329\_1330, located on St.\ Petersburger Street, recording reductions of 6,959 and 5,592 vehicles per day, respectively. These substantial decreases suggest that traffic formerly using the Carola Bridge no longer accesses the city centre via this corridor, instead being rerouted to alternative crossings and corridors. In contrast, Albert Bridge-related arteries record absolute increases ranging from +1,314 to +4,328 vehicles per day, supporting the bridge’s new central role as the popular route for river crossing. Marien Bridge-related arteries also display upward trends, with increases between +1,267 and +4,844 vehicles per day.

\begin{table}[h!]
\centering
\renewcommand\arraystretch{0.91}
\captionsetup{font={small}}
\footnotesize
\begin{center}
  \caption{Daily traffic flow changes on critical roads. The details of the detector IDs are given in Table \ref{tab:critical-detectors-compact}}
  \vspace{2ex}
  \label{flow change}
  \begin{tabular}{%
    >{\raggedright\arraybackslash}p{3.2cm}
    >{\centering\arraybackslash}p{1.2cm}
    >{\centering\arraybackslash}p{1.3cm}
    >{\raggedleft\arraybackslash}p{1.5cm}
    >{\raggedleft\arraybackslash}p{1.5cm}
    >{\raggedleft\arraybackslash}p{1.5cm} 
    >{\raggedleft\arraybackslash}p{1.5cm}
  }
  \hline
  Critical roads & Direction & Detector index & \makecell{Traffic\\volume\\before\\collapse\\(Veh/day)} & \makecell{Traffic\\volume\\after\\collapse\\(Veh/day)} & \makecell{Volume \\change\\(Veh/day)} & \makecell{Change\\percentage} \\
  \hline
  \multirow{2}{=}{Marien Bridge} & outbound & 1 & 13818 & 17164 & +3345 & +24.20\% \\
                                 & inbound  & 2 & 15004 & 17742 & +2738 & +18.25\% \\
  \hline
  \multirow{2}{=}{Albert Bridge} & outbound & 3 & 8838 & 15688 & +6849 & +77.49\% \\
                                 & inbound  & 4 & 8671 & 15720 & +7048 & +81.28\% \\
  \hline
  \multirow{4}{=}{Outer New City--Inner New City Corridor} &  &  &  &  &  & \\
                                                           & outbound & 5 & 7906 & 1987 & --5918 & --74.86\% \\
                                                           & inbound  & 6 & 7584 & 2193 & --5390 & --71.08\% \\
                                                           &  &  &  &  &  & \\
                                                           
  \hline
  \multirow{4}{=}{Leipzig Suburb--Inner New City Corridor} & \multirow{1}{=}{outbound} & 7 & 14301 & 12286 & --2014 & --14.08\% \\
                                                           & \multirow{3}{=}{inbound} & 8 & 6192 & 2836 & --1504 & --16.95\% \\
                                                           &                         & 9 & 12324 & 7942 & --4382 & --35.56\% \\
                                                           &                         & 10 & 12306 & 7820 & --4485 & --36.45\% \\
  \hline
  \multirow{4}{=}{Inner New City East--West Corridor} & \multirow{2}{=}{outbound} & 11 & 1384 & 1655 & +271 & --19.60\% \\
                                                      &                         & 12 & 5122 & 8061 & +2938 & +57.37\% \\
                                                      & \multirow{2}{=}{inbound} & 13 & 2172 & 2554 & +381 & +17.55\% \\
                                                      &                         & 14 & 5071 & 8617 & +3546 & +69.93\% \\
                                                      
  \hline
  \multirow{7}{=}{Marien Bridge-related artery} & \multirow{3}{=}{inbound} & 15 & 11856 & 13124 & +1267 & +10.69\% \\
                                                &                           & 16 & 9756 & 11597 & +1840 & +18.87\% \\
                                                &                           & 17 & 13038 & 15852 & +2813 & +21.58\% \\
                                                & \multirow{4}{=}{outbound} & 18 & 8893 & 11881 & +2988 & +33.60\% \\
                                                &                           & 19 & 8783 & 11115 & +2332 & +26.55\% \\
                                                &                           & 20 & 13574 & 18419 & +4844 & +35.69\% \\
                                                &                           & 21 & 2834 & 4458 & +1623 & +57.30\% \\
  \hline
  \multirow{5}{=}{Carola Bridge-related artery} & \multirow{3}{=}{inbound} & 22 & 15410 & 8451 & --6959 & --45.16\% \\
                                                &                           & 23 & 18157 & 13388 & --4768 & --26.26\% \\
                                                &                           & 24 & 11334 & 9235 & --2099 & --18.52\% \\
                                                & \multirow{2}{=}{outbound} & 25 & 10006 & 7241 & --2764 & --27.63\% \\
                                                &                           & 26 & 16871 & 11278 & --5592 & --33.15\% \\
  \hline
  \multirow{5}{=}{Albert Bridge-related artery} & \multirow{3}{=}{inbound} & 27 & 3156 & 5274 & +2108& +66.61\% \\
                                                &                           & 28 & 6653 & 10053 & +3400 & +51.11\% \\
                                                &                           & 29 & 5786 & 7988 & +2201 & +38.05\% \\
                                                & \multirow{2}{=}{outbound} & 30 & 5532 & 6847 & +1314 & +23.75\% \\
                                                &                           & 31 & 6021 & 10350 & +4328 & +71.90\% \\
  \hline
  \end{tabular}
\end{center}
\end{table}

\subsection{Statistical Validation}

    To determine if the observed changes in traffic volume are statistically significant, we perform paired T-tests and one-way ANOVA tests. Table \ref{tab:stat_summary} summarises these results for key arterial roads and corridors. The analysis shows that the changes on all critical road segments are highly statistically significant. The T-statistics and F-statistics are universally significant at the $p < 0.01$ level, confirming that the observed shifts are a direct consequence of the collapse and not a result of random daily fluctuations. The t-statistics illustrate the direction and magnitude of the traffic redistribution. For instance, the large negative t-statistics for the Albert Bridge (e.g., –56.80 for outbound traffic) and its related arteries confirm a massive and statistically significant increase in traffic. Conversely, the large positive t-statistics on Carola Bridge-related arteries (e.g., +49.09) and the Outer New City–Inner New City Corridor (+62.79) demonstrate a significant decrease in volume as traffic diverts away from these routes.

In addition, we calculate the EFCH value, defined as the change in average hourly traffic flow normalised by the maximum observed operational flow (95th percentile), to quantify the relative stress on each road segment. The Albert Bridge experiences the highest saturation, with EFCH value exceeding 3.4 in both directions, indicating that post-collapse traffic volumes exceed three times the typical high-flow operating threshold. Similarly, the EFCH value of arteries connected to the Marien Bridge range from 1.8 to 3.7. In contrast, the corridors previously linked to the Carola Bridge experience traffic declines, evidenced by negative EFCH value. Segments such as 1310\_1311 and 1329\_1330 on St. Petersburger Street exhibit EFCH value below –2.5, suggesting substantial underutilisation or loss of functionality after the collapse.

Beyond mean shifts, the F-statistics add a dynamic dimension by measuring variability between the pre- and post-collapse periods. Extremely high values, for instance $F = 2245$ on segment 1371 and $F = 1874$ on 1322\_1323 on Albert street, indicate sharp fluctuations and temporal instability. Similar patterns are observed on Albert Bridge-related arteries, pointing to irregular daily demand. Meanwhile, segments with moderate F-statistics (for example, those associated with the Marien Bridge) indicate relatively stable adaptation.

Moreover, additional insights from the statistical analysis that are not evident from the volume differences highlight directional asymmetry in traffic adaptation. For example, on the Marien Bridge-related arteries, outbound EFCHs (e.g., 3.76 on 1416) are much higher than inbound ones (e.g., 1.43 on 911\_912), suggesting stronger pressure in the outbound direction, possibly linked to morning commuting peaks or route preferences. A similar asymmetry is evident in the Inner New City East–West Corridor, where inbound segment 1361\_1360 shows stronger saturation (EFCH = 1.80) compared with 1239 (EFCH = 0.29), despite similar relative increases in traffic flow.

\begin{table}[h!]
\centering
\renewcommand\arraystretch{0.91}
\captionsetup{font={small}}
\footnotesize
\begin{center}
  \caption{Summary of statistical tests on critical roads. The details of the detector IDs are given in Table \ref{tab:critical-detectors-compact}}
  \vspace{2ex}
  \label{tab:stat_summary}
  \begin{tabular}{%
    >{\raggedright\arraybackslash}p{3.5cm}
    >{\centering\arraybackslash}p{1.2cm}
    >{\centering\arraybackslash}p{1.8cm}
    >{\raggedleft\arraybackslash}p{1.5cm}
    >{\raggedleft\arraybackslash}p{1.5cm}
    >{\raggedleft\arraybackslash}p{1.5cm} 
    >{\raggedleft\arraybackslash}p{1.5cm}
  }
  \hline
  Critical roads & Direction & Detector index & T-statistic & F-statistics &  Equivalent Full-Capacity Hours \\
  \hline
  \multirow{2}{=}{Marien Bridge} & outbound & 1 & $-25.82$ & $106.61$ & $1.29$ \\
                                 & inbound  & 2 & $-28.56$ & $63.45$ & $1.39$ \\
  \hline
  \multirow{2}{=}{Albert Bridge} & outbound & 3 & $-56.79$ & $553.79$ & $3.48$  \\
                                 & inbound  & 4 & $-50.26$ & $481.89$ & $3.58$  \\
  \hline
  \multirow{3}{=}{Outer New City--Inner New City Corridor} & outbound & 5 & $62.78$ & $2245$ & $-4.59$ \\
                                                           & inbound  & 6 & $60.50$ & $1874$ & $-2.74$  \\
                                                           &  &  &  &  &  \\
                                                           
  \hline
  \multirow{4}{=}{Leipzig Suburb--Inner New City Corridor} & \multirow{1}{=}{outbound} & 7 & $23.30$ & $48.78$  & $-1.02$  \\
                                                           & \multirow{3}{=}{inbound} & 8 & $51.02$ & $875.71$ & $-1.70$ \\
                                                           &                         & 9 & $45.22$ & $373.83$ & $-2.23$\\
                                                           &                         & 10 & $41.92$ & $395.06$ & $-2.28$  \\
  \hline

  \multirow{5}{=}{Inner New City East--West Corridor} & \multirow{2}{=}{outbound} & 11 & $-15.51$ & $43.04$ & $0.21$  \\
                                                      &                         & 12 & $-23.91$ & $168.90$ & $1.50$  \\
                                                      & \multirow{3}{=}{inbound} & 13 & $-14.61$ & $37.55$ & $0.29$ \\
                                                      &                         & 14 & $-54.22$ & $513.66$ & $1.80$  \\
                                                      &  &  &  &  &  \\
                                                      
  \hline

  \hline

  \multirow{7}{=}{Marien Bridge-related artery} & \multirow{3}{=}{inbound} & 15 & $-11.42$ & $21.31$ & $0.64$  \\
                                                &                           & 16 & $-28.43$ & $64.87$ & $1.42$  \\
                                                &                           & 17 & $-36.53$ & $75.25$ & $1.43$  \\
                                                & \multirow{4}{=}{outbound} & 18 & $-42.10$ & $159.84$ & $1.52$ \\
                                                &                           & 19 & $-31.68$ & $116.39$ & $1.18$  \\
                                                &                           & 20 & $-44.60$ & $196.91$ & $3.76$ \\
                                                &                           & 21 & $-33.97$ & $348.39$ & $1.26$  \\
  \hline
  \multirow{5}{=}{Carola Bridge-related artery} & \multirow{3}{=}{inbound} & 22 & $9.61$ & $327.24$ & $-3.54$  \\
                                                &                           & 23 & $41.16$ & $157.19$ & $-2.42$ \\
                                                &                           & 24 & $27.20$ & $68.36$ & $-1.06$ \\
                                                & \multirow{2}{=}{outbound} & 25 & $37.69$ & $188.65$ & $-1.14$  \\
                                                &                           & 26 & $49.09$ & $292.59$ & $-2.84$  \\
  \hline
  \multirow{5}{=}{Albert Bridge-related artery} & \multirow{3}{=}{inbound} & 27 & $-35.75$ & $338.48$ & $1.63$ \\
                                                &                           & 28 & $-43.89$ & $270.49$ & $1.73$  \\
                                                &                           & 29 & $-35.06$ & $138.85$ & $1.70$  \\
                                                & \multirow{2}{=}{outbound} & 30 & $-30.46$ & $73.17$ & $1.02$  \\
                                                &                           & 31 & $-52.41$ & $524.85$ & $3.36$  \\
  \hline
  \end{tabular}
\end{center}
\end{table}

\begin{figure}
    \centering
    \includegraphics[width=1\linewidth]{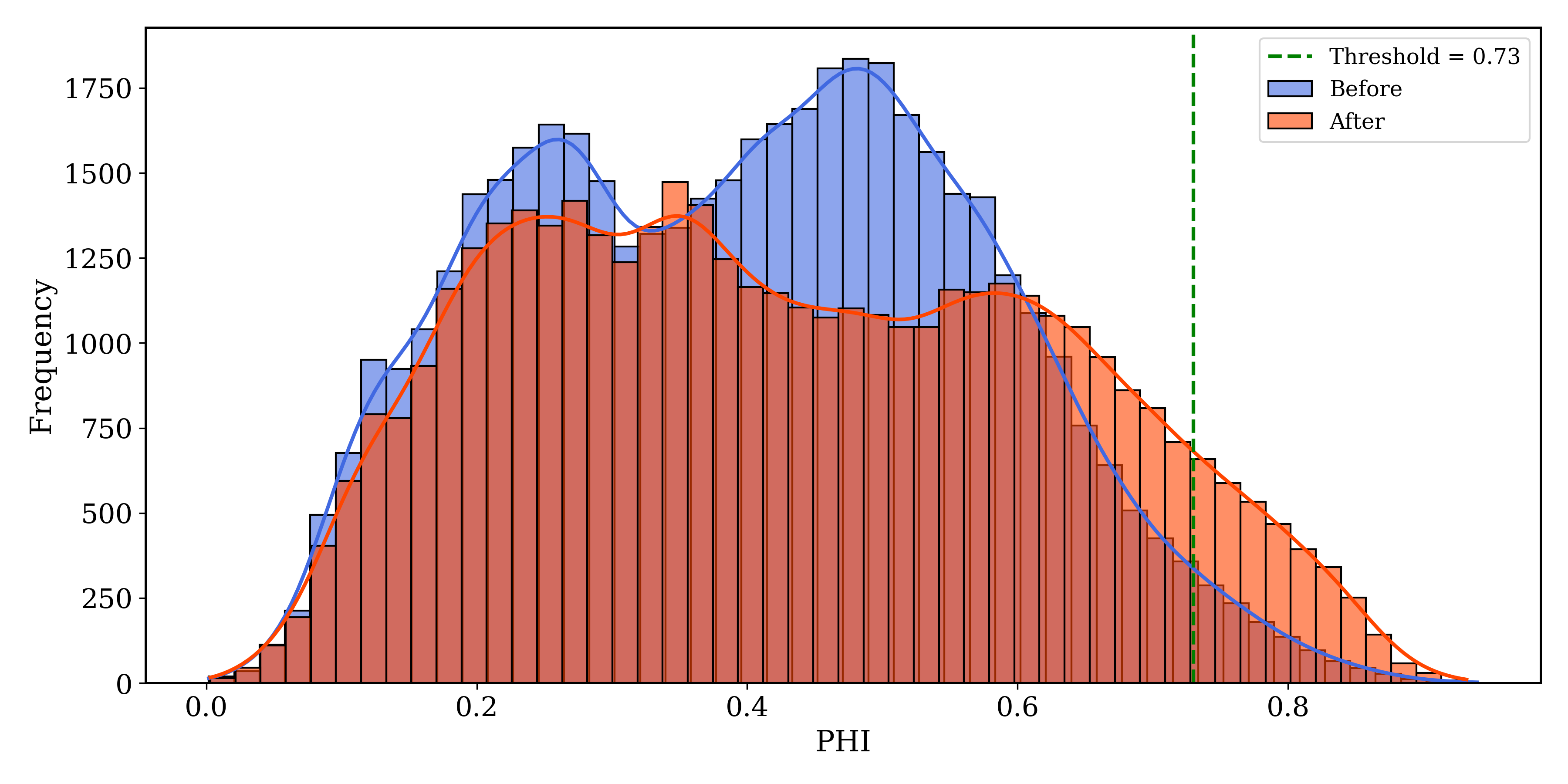}
    \caption{Distribution of the PHI values before and after the Carola Bridge collapse. The blue histogram represents PHI values for the before days, while the red histogram shows the values of the after days. The vertical dashed green line indicates the derived PHI threshold (0.73), used to distinguish the peak period.}
    \label{fig:phi_distribution}
\end{figure}

\subsection{Analysis of Peak Hour}
To comprehensively understand the temporal shift in daily travel patterns following the collapse, we perform a detailed analysis using the proposed PHI. Figure~\ref{fig:phi_distribution} illustrates the distribution of PHI values for the periods before and after the collapse. The threshold is calculated using the aforementioned formula (\ref{eq2}) and determined to be 0.73. Before the collapse, as shown by the blue histogram, the PHI distribution exhibits a distinct bimodal pattern. One peak is centred at approximately 0.28, with a larger second peak around 0.50. This bimodal nature suggests that prior to the collapse, the network operates under normal conditions with regular peak hours. Few samples exceed the threshold, indicating that peak-hour conditions are not prolonged.

After the collapse, the distribution’s mass noticeably shifts to the right, indicating a general increase in PHI values, and the curve becomes more unimodal. This transformation in the PHI distribution provides clear evidence of the bridge collapse’s impact, showing that the system transitions from a stable state with predictable bimodal behaviour to a more volatile and less predictable one. This shift indicates that numerous road segments experience extended peak hours. Therefore, we further investigate the changes in these critical road segments using heatmaps.

\begin{figure}[!t]
    \centering
    \begin{subfigure}[b]{0.48\textwidth}
        \centering
        \includegraphics[width=\textwidth]{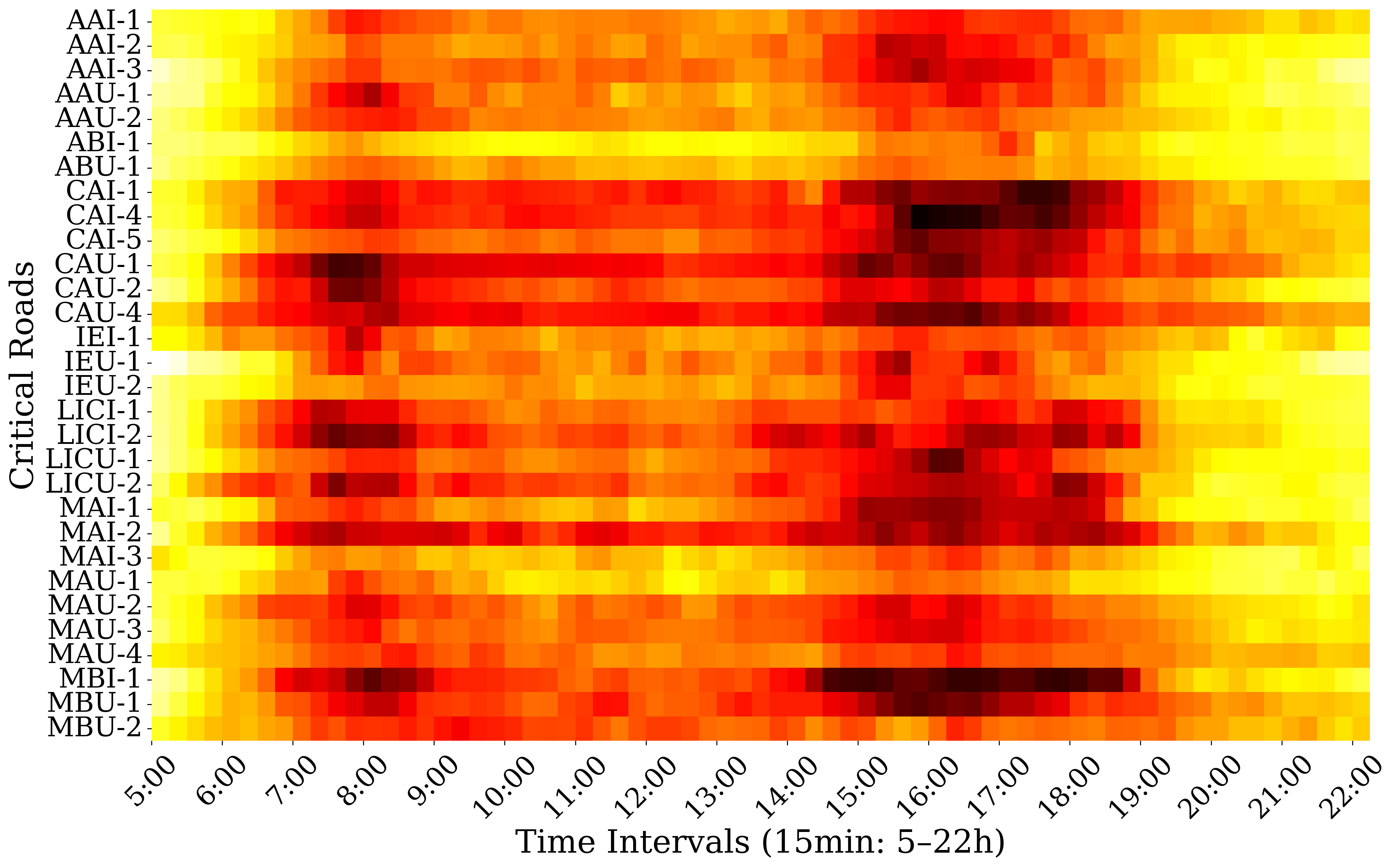}
        \caption{Monday before}
        \label{fig:mon_before}
    \end{subfigure}
    \begin{subfigure}[b]{0.48\textwidth}
        \centering
        \includegraphics[width=\textwidth]{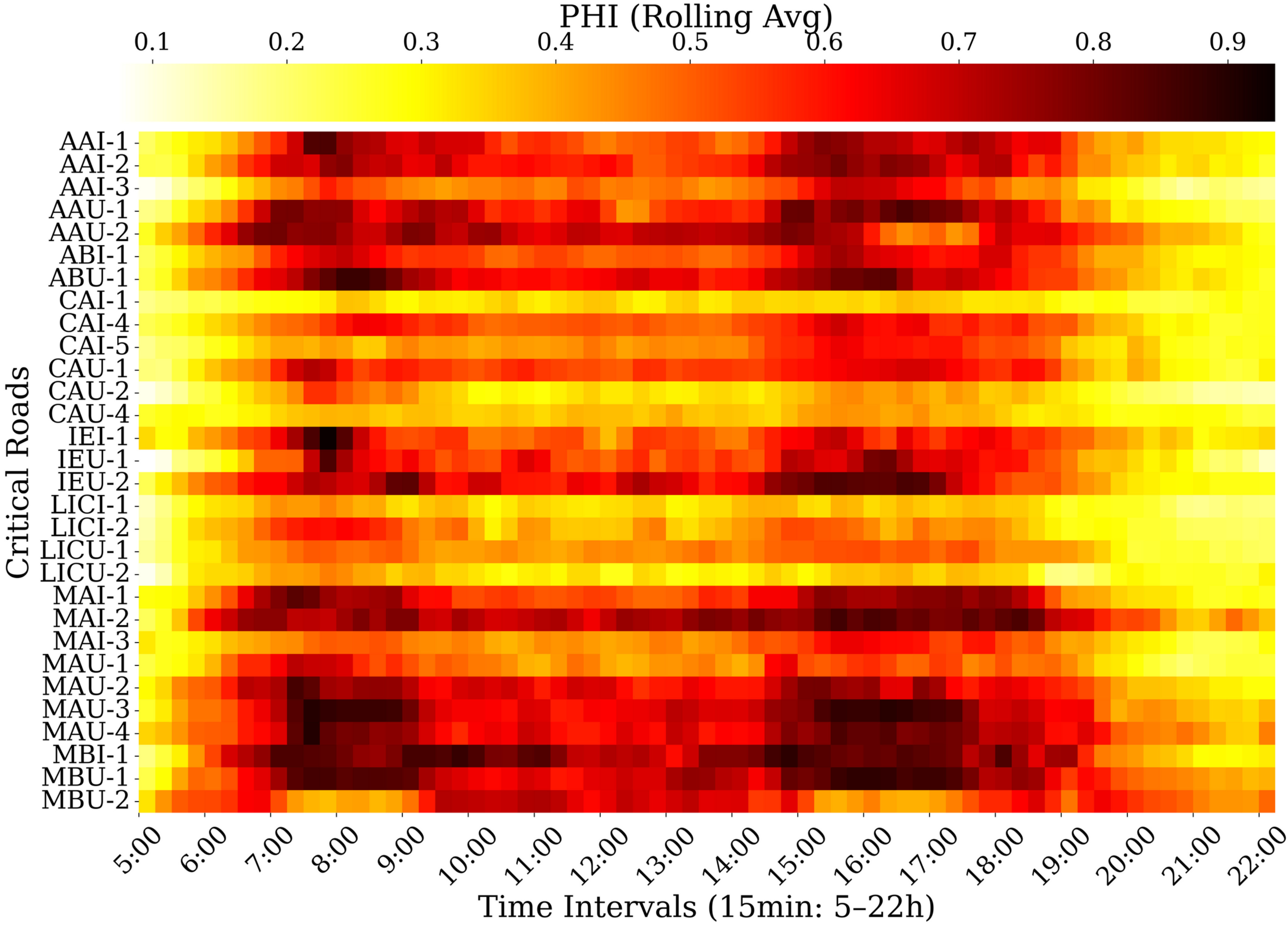}
        \caption{Monday after}
        \label{fig:mon_after}
    \end{subfigure}
    \centering
    \begin{subfigure}[b]{0.48\textwidth}
        \centering
        \includegraphics[width=\textwidth]{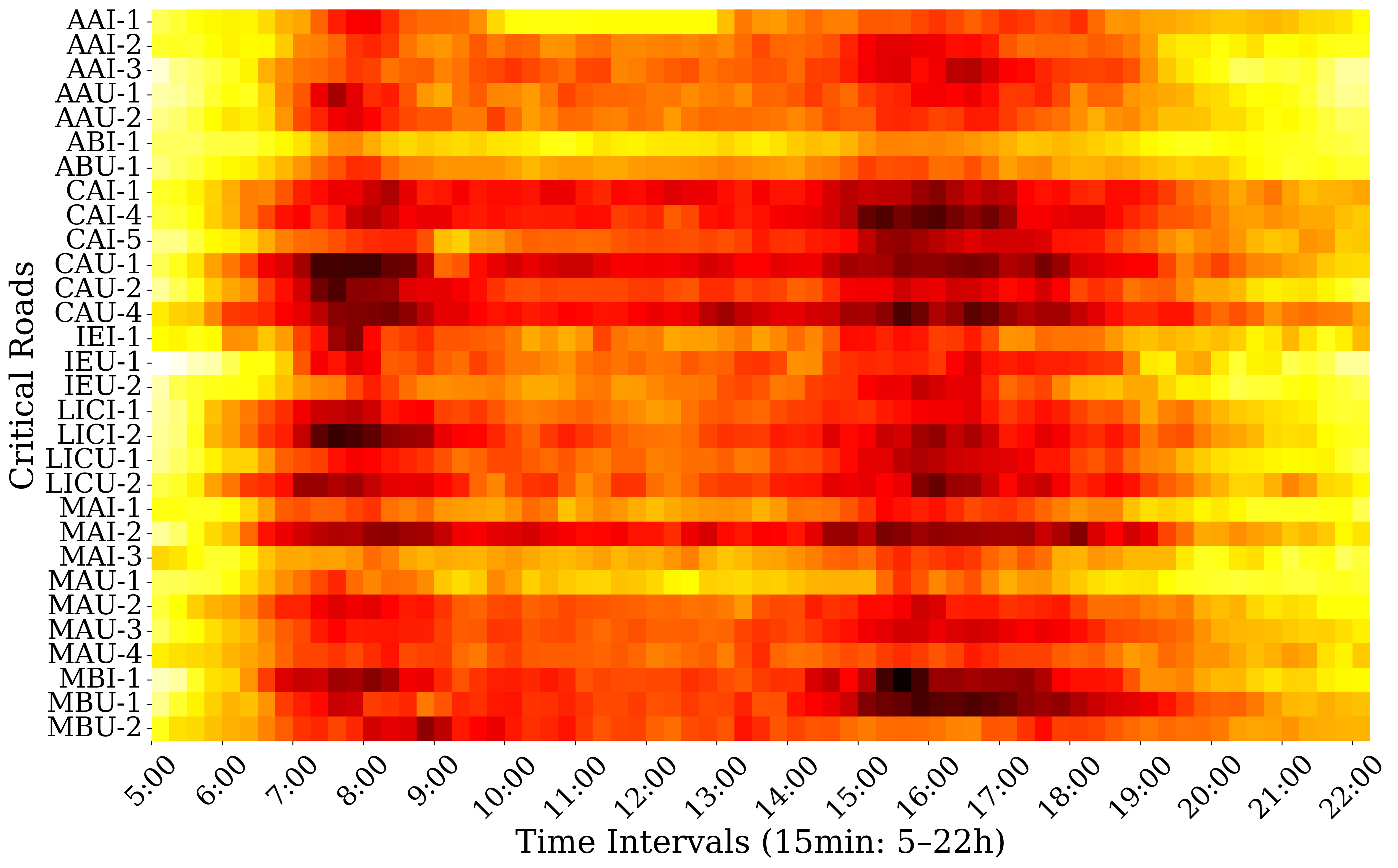}
        \caption{Wednesday before.}
        \label{fig:wed_before}
    \end{subfigure}
    \begin{subfigure}[b]{0.48\textwidth}
        \centering
        \includegraphics[width=\textwidth]{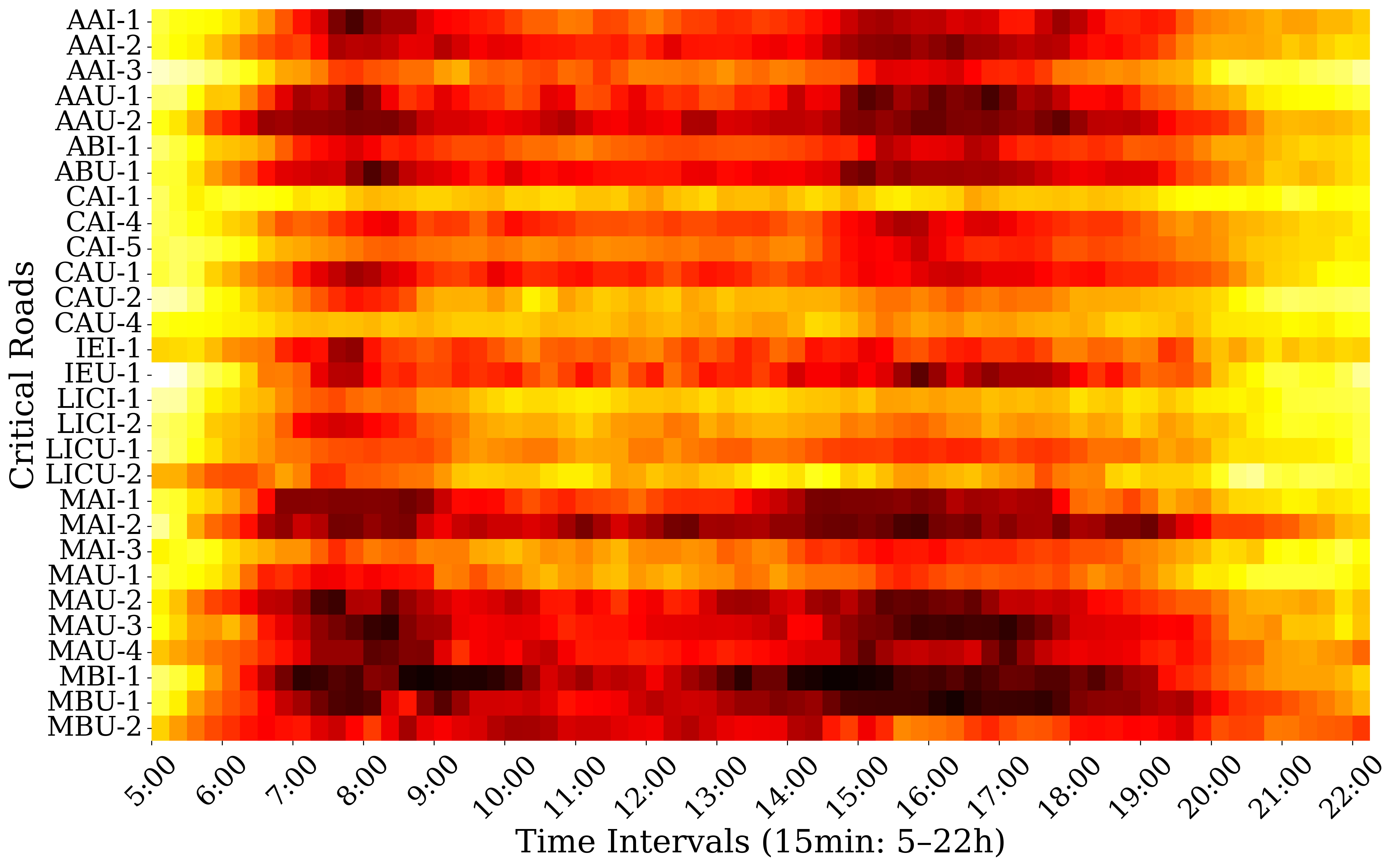}
        \caption{Wednesday after}
        \label{fig:wed_after}
    \end{subfigure}
    \centering
    \begin{subfigure}[b]{0.48\textwidth}
        \centering
        \includegraphics[width=\textwidth]{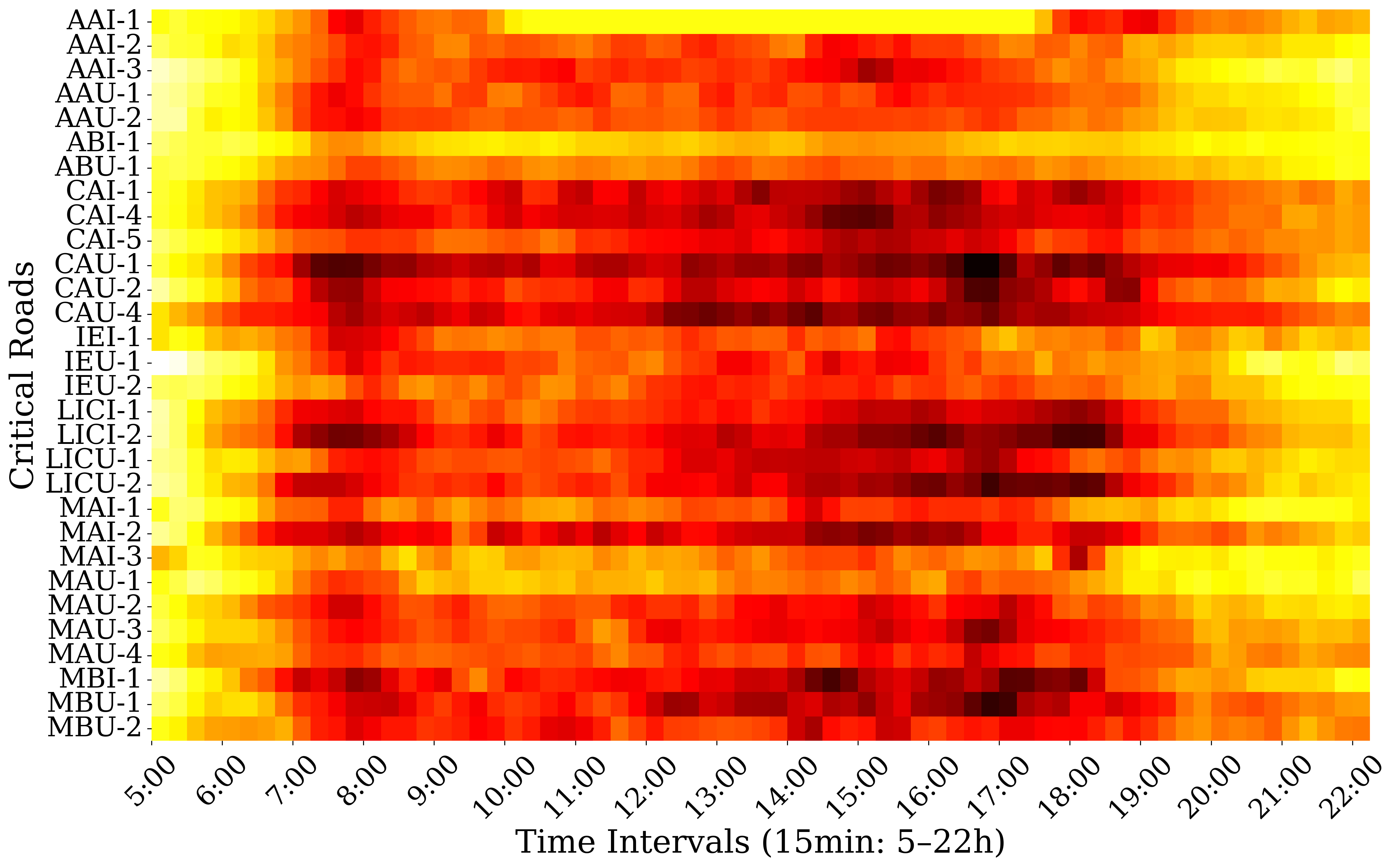}
        \caption{Friday before.}
        \label{fig:fri_before}
    \end{subfigure}
    \begin{subfigure}[b]{0.48\textwidth}
        \centering
        \includegraphics[width=\textwidth]{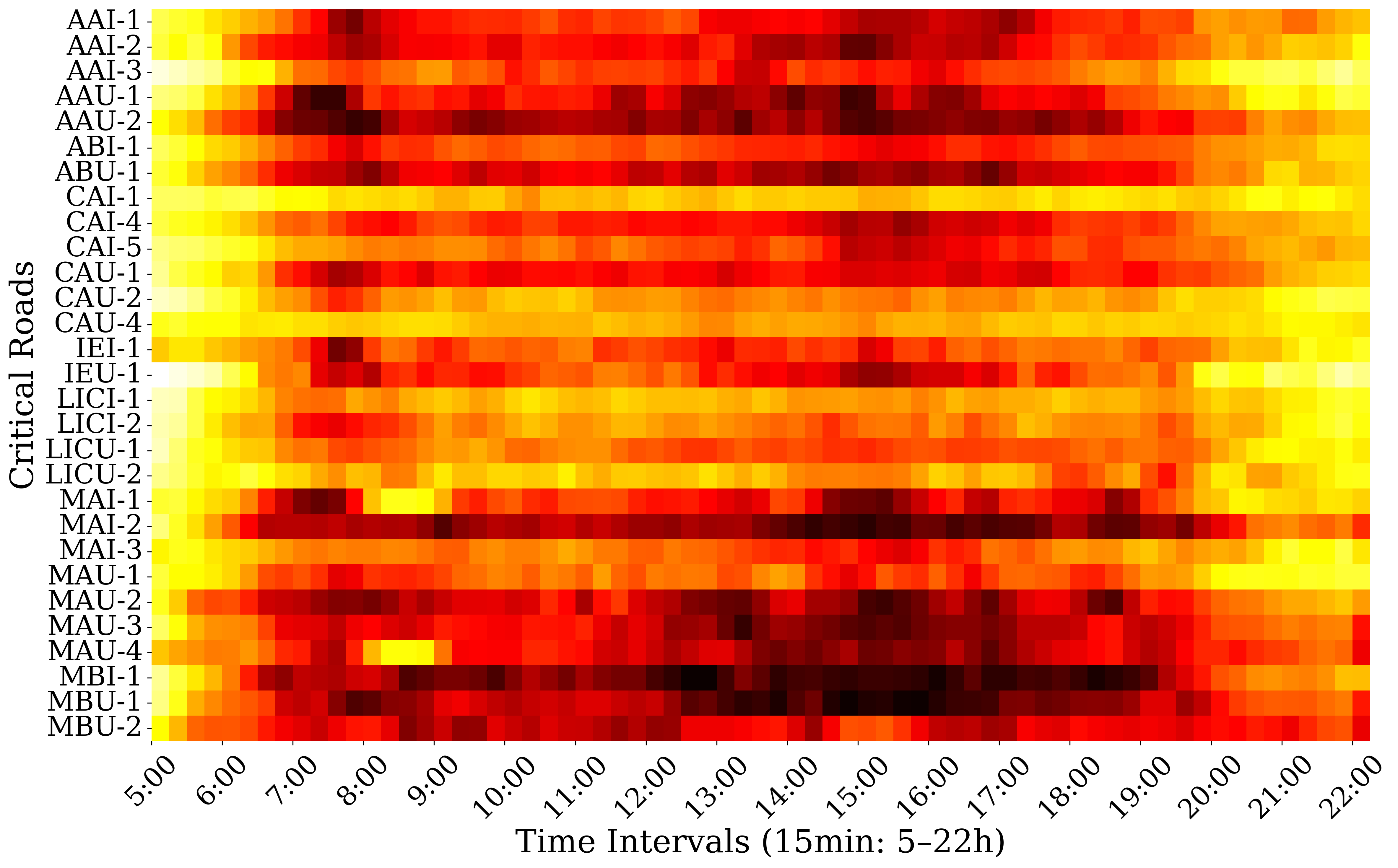}
        \caption{Friday after}
        \label{fig:fri_after}
    \end{subfigure}
    \caption{Analysis of daily peak hour changes after the Carola bridge collapse. Heatmaps represent the PHI values across critical roads. For the other days, the heatmaps are provided in Appendix \ref{fig:phi_comparison_app}. The details of the detectors and their locations are provided in Table \ref{tab:critical-detectors-compact} and can be matched with the labels shown in the figure.}
    \label{fig:phi_comparison}
\end{figure}

\begin{figure}[!t]
    \centering
    \begin{subfigure}[b]{0.7\textwidth}
        \centering
        \includegraphics[width=\textwidth]{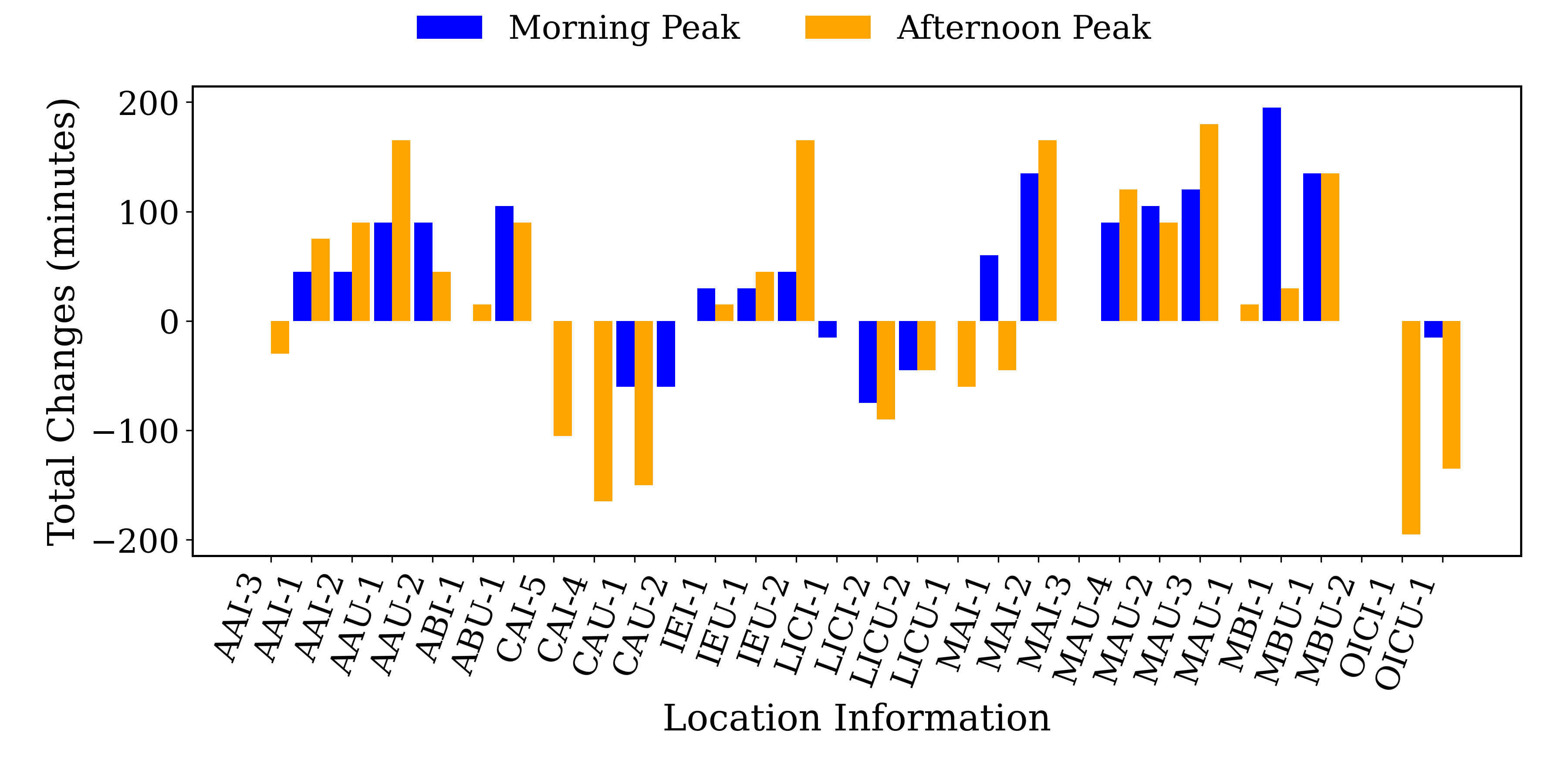}
        \caption{Monday Minute Changes.}
        \label{fig:minutes_changes_mon}
    \end{subfigure}
    \begin{subfigure}[b]{0.7\textwidth}
        \centering
        \includegraphics[width=\textwidth]{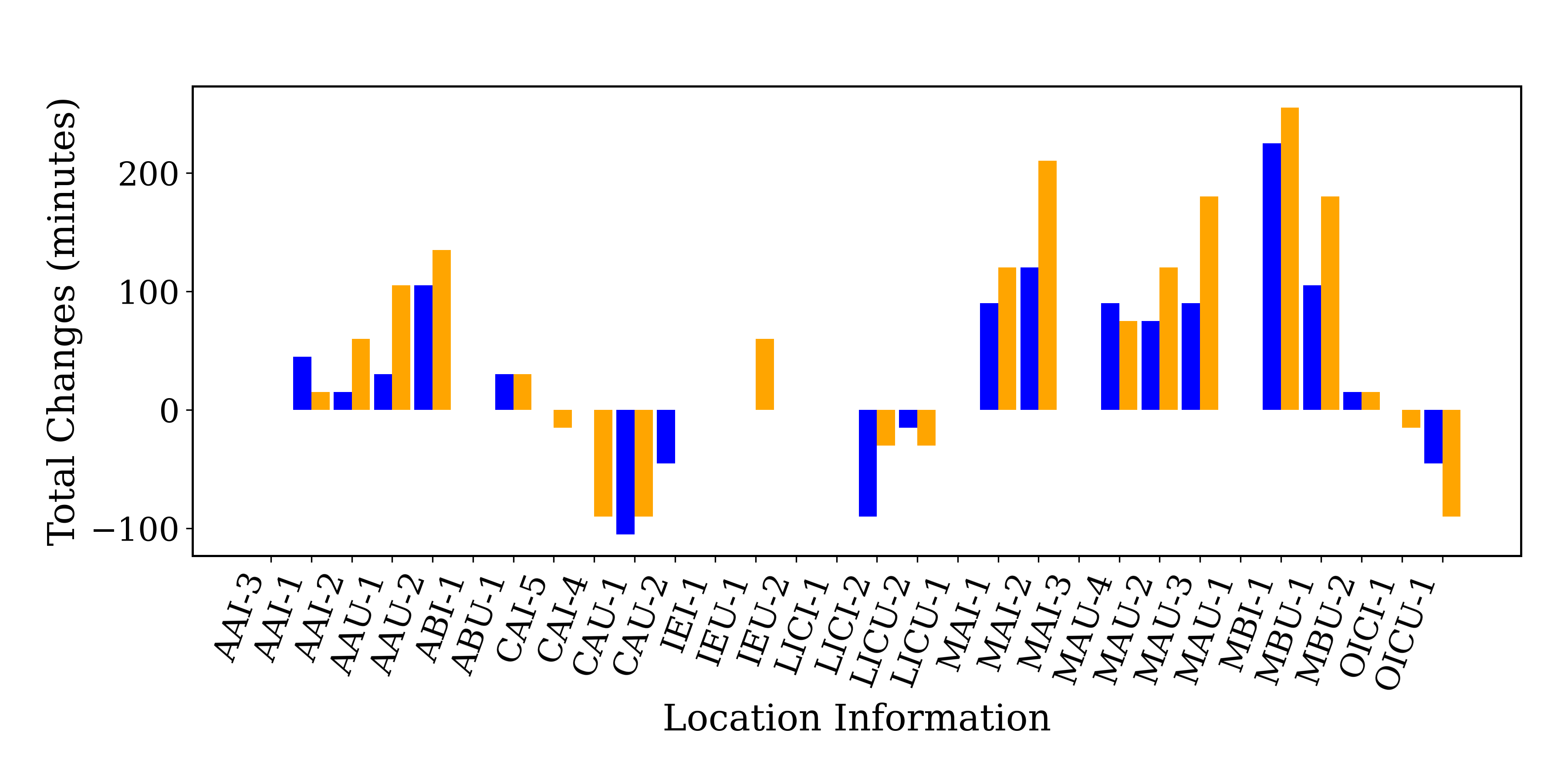}
        \caption{Wednesday Minute Chnages}
        \label{fig:minutes_changes_wed}
    \end{subfigure}
    \begin{subfigure}[b]{0.7\textwidth}
        \centering
        \includegraphics[width=\textwidth]{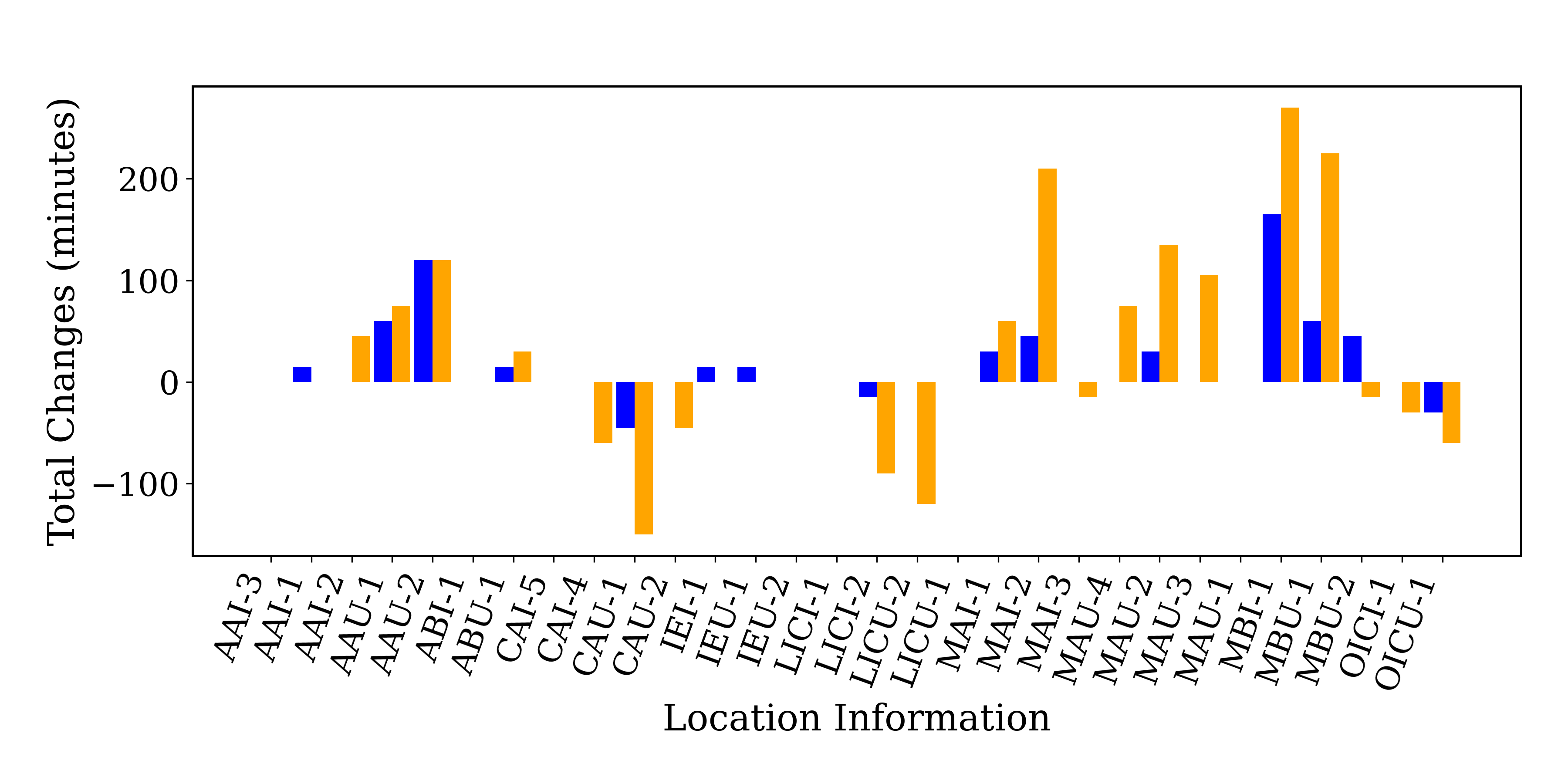}
        \caption{Friday Minute Chnages}
        \label{fig:minutes_changes_fri}
    \end{subfigure}
    \caption{Changes in morning and afternoon peak durations (in minutes) across the critical road on three weekdays. For the other days, the plots are provided in Appendix \ref{fig:minute_chnages_app}. The details of the detectors and their locations are provided in Table \ref{tab:critical-detectors-compact} and can be matched with the labels shown in the figure.}
    \label{fig:minute_changes}
\end{figure}

Figures \ref{fig:phi_comparison} present the PHI heatmaps for the critical road segments before and after the collapse for different weekdays, respectively. Besides, the total morning and afternoon peak hour minute changes are presented in Figure \ref{fig:minute_changes}. Morning peak changes are calculated using values up to 12:00 PM, whereas afternoon peak changes are calculated from values after 12:00 PM. It is clearly visible that peak hours extend significantly compared with the pre-collapse period, particularly on the Albert Bridge and its associated arterial roads. The total morning peak-hour changes in this artery range from approximately 50 to 100 minutes, and from 50 to 150 minutes for the afternoon peak hours. This finding is further supported by the quantitative peak-hour shift analysis. Prior to the collapse, the Carola Bridge and its connecting roads are among the busiest; however, after the collapse, a noticeable drop in PHI is observed along these routes, indicating that travellers begin avoiding them. Interestingly, the most substantial extension in peak hours is observed on the Marien Bridge. Despite this, our analysis shows that the majority of the diverted traffic shifts toward the Albert Bridge, which experiences nearly four times the increase in vehicle volume compared with the Marien Bridge. The reason for the more severe congestion on the Marien Bridge lies in its pre-existing condition—it already operates at or near capacity before the collapse. Therefore, even a moderate increase in volume leads to disproportionately higher congestion and extended peak periods. For the Marien Bridge and its related arteries, the morning peak-hour changes range from around 50 to 200 minutes, and similarly, the afternoon peak-hour changes extend up to 250 minutes. At most alternative route locations, both morning and afternoon peak hours expand. This shift likely reflects changes in travel behaviour, as commuters begin departing earlier or later than usual to avoid congestion. The PHI values offer a subtle perspective on how network disruptions influence travel dynamics. While the absolute change in PHI during peak and off-peak hours is similar, the relative increase is more pronounced during off-peak periods. This indicates that many travellers respond to the disruption by adjusting their travel times and route choices more flexibly, possibly using two alternative routes to travel in and out of the city.

In addition, we also investigate how the peak hour changes across different weekdays. We find that almost all weekdays experience noticeable peak-hour extensions. However, Friday (Figure~\ref{fig:fri_before} and Figure~\ref{fig:fri_after}) shows comparatively smaller changes than the other weekdays. This might be attributed to the nature of tourist travel or people heading out of the city. As visible in the figure, the morning peak hour on Friday changes only slightly, indicating that regular commuting behaviour does not significantly alter. On the other hand, the evening peak hour is more extended, which might be due to incoming tourist traffic or people arriving in Dresden for the weekend. Interestingly, on Monday (Figure~\ref{fig:mon_before} and Figure~\ref{fig:mon_after}), we observe more significant changes in the morning peak hour, especially in locations such as the Marien Bridge and Albert Bridge. This suggests that people are likely leaving the city after the weekend. Since Dresden is a well-known tourist destination, this aligns with the assumption that many visitors plan their return trips on Monday morning to avoid weekend congestion \citep{Fina2020}.

From the empirical analysis of the network-wide spatio-temporal effects, we gain an understanding of how traffic patterns change both in total and on a daily basis. We also examine how these changes affect within-day traffic patterns, finding extended peak-hour periods on certain roads as well as differing behaviours between inbound and outbound travel. Therefore, in the following section, we investigate commuter behavioural adaptations, including possible modal shifts and the use of alternative routes.

\subsection{Analysis of Commuter Behavioural Adaptation}

\begin{figure}[!t]
    \centering
    \includegraphics[width=1\linewidth]{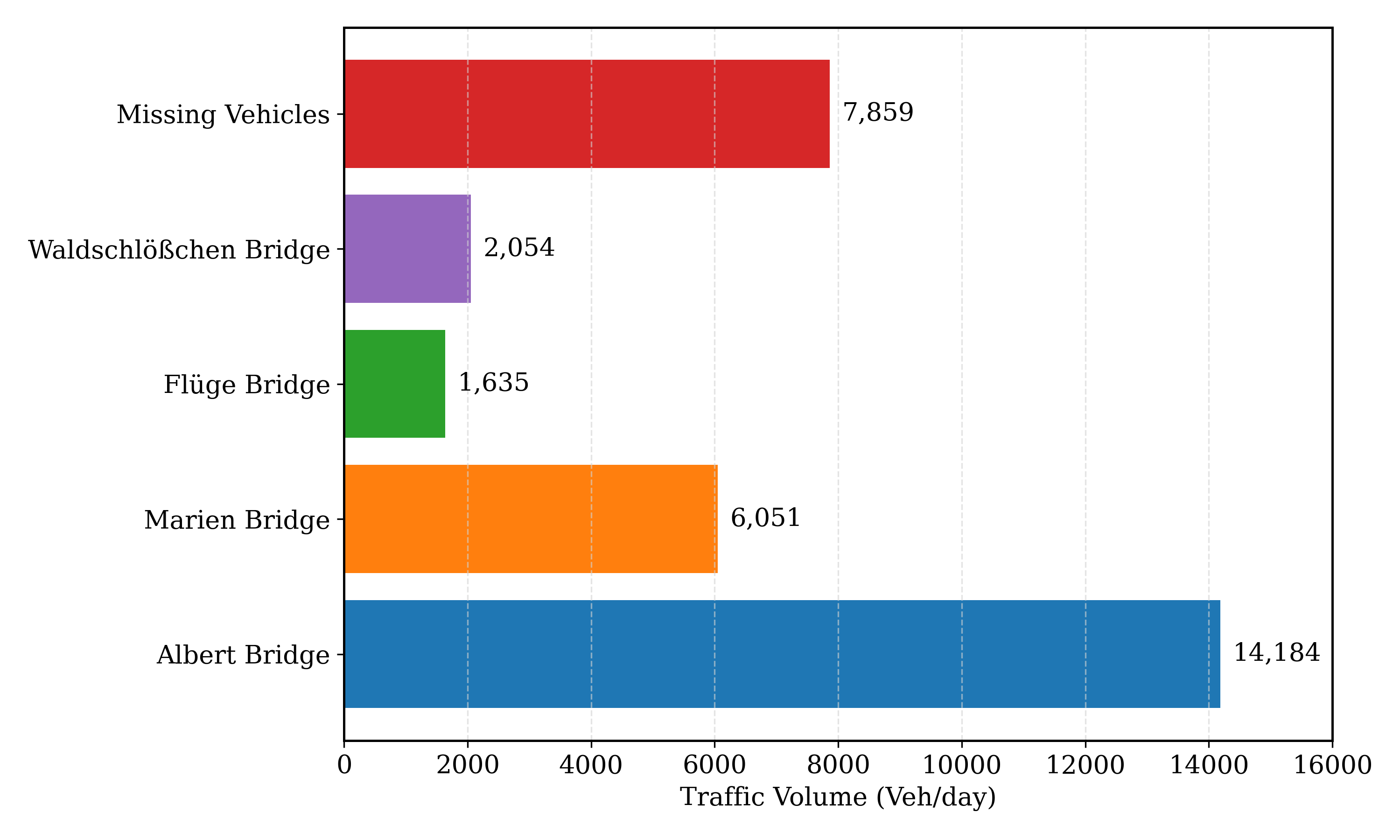}
    \caption{Redistribution of daily traffic volume from Carola Bridge to the alternative river crossing bridges following the collapse.}
    \label{fig:reditribution}
\end{figure}

Before the collapse, the Carola Bridge serves approximately 30,000 to 32,000 vehicles per day, with seasonal fluctuations. Figure~\ref{fig:reditribution} illustrates the redistribution of this traffic volume across alternative bridges following the collapse. The analysis shows that the largest share of diverted traffic, approximately 14,184 vehicles per day, shifts to the Albert Bridge, accounting for nearly 45\% of the original Carola Bridge volume. The Marien Bridge attracts around 6,051 vehicles per day, roughly a quarter of the redistributed traffic. Together, these two bridges carry more than two-thirds of the diverted flow. In contrast, the Waldschlößchen Bridge records a modest increase of 2,054 vehicles per day (around 6\%), while the Flügelweg Bridge absorbs 1,635 vehicles per day (about 5\%). Notably, approximately 8,000 vehicles per day remain unaccounted for (about 25\% of the Carola Bridge’s prior demand), suggesting a reduction in car travel demand. To translate this into passenger terms, we assume that private cars account for around 90\% of traffic in Dresden’s city centre. Based on this and an average vehicle occupancy of 1.3 persons per car ~\citep{europaOccupancyRates}, the missing vehicles corresponds to nearly 9,180 unaccounted passengers per day. These missing trips likely reflect a combination of factors such as modal shifts to other modes of transport, increased reliance on remote work, or re-routing along corridors outside the VAMOS monitoring network.

\begin{figure}[!t]
    \centering
    \includegraphics[width=1\linewidth]{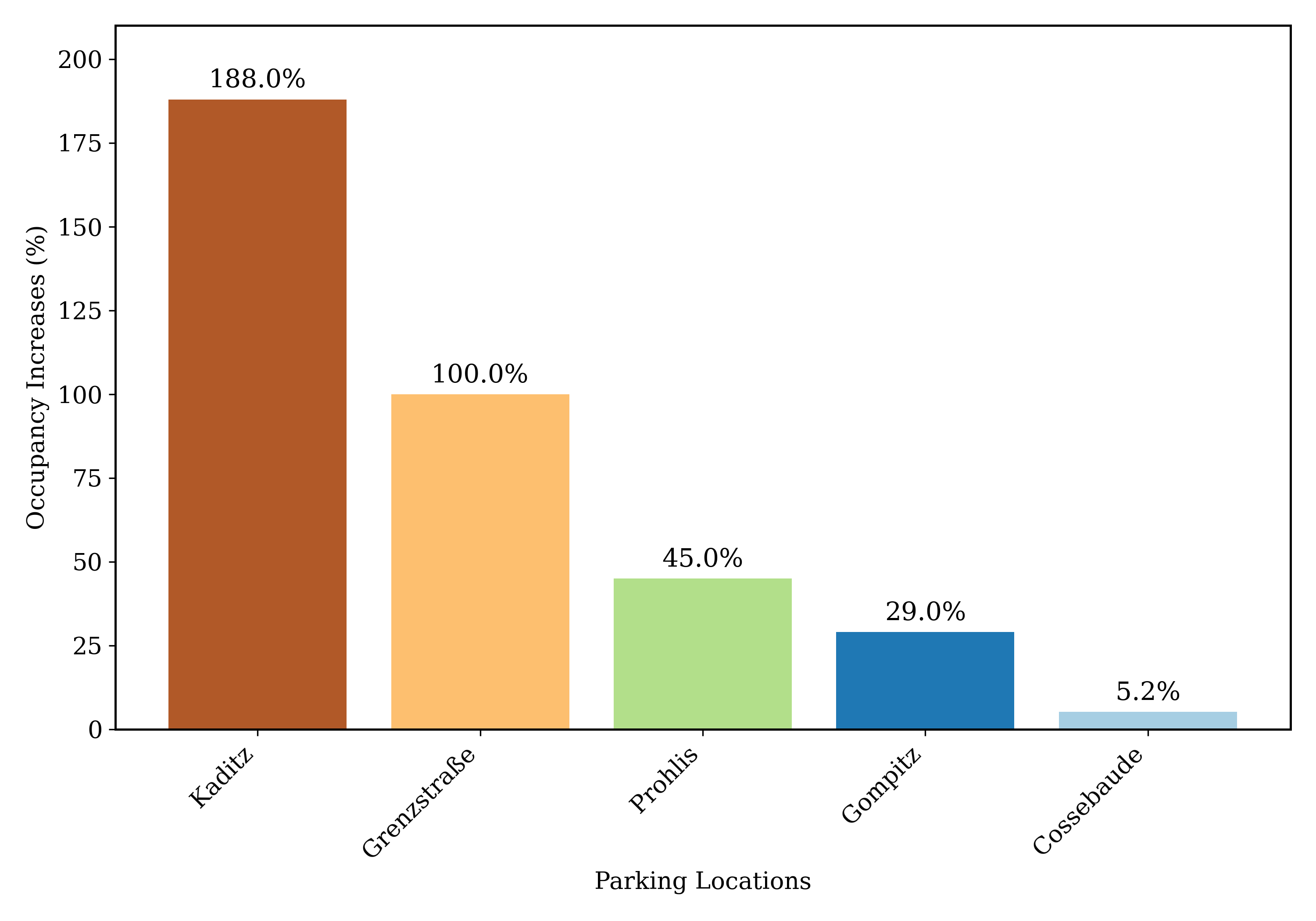}
    \caption{Comparison of occupancy of P+R parking}
    \label{fig:pr_parking}
\end{figure}

As indicated in prior analyses, the collapse of the Carola Bridge may have triggered modal shifts, decreased car traffic demands, or changes in route choice. To explore these behavioural adaptations, we investigate P+R usage data before and after the collapse to assess whether long-distance commuters shifted to public transport. Dresden’s P+R system is uniquely incentivized, drivers can park free of charge if they continue their journey by public transport. Figure \ref{fig:pr_parking} illustrates a significant rise in P+R usage across several facilities following the collapse, with Kaditz and Grenzstraße experiencing increases of 188\% and 100\%, respectively. This surge suggests that commuters increasingly opted for multimodal travel—likely driven by a desire to avoid inner-city congestion and reduce overall travel time. This behavioural change may also be supported by Germany’s nationwide “Deutschlandticket,” which offers low-cost public transit access and additional subsidies for job commuters, making public transport more attractive, especially for suburban travellers. Although this study did not evaluate bicycle usage, it is plausible that some commuters also shifted to cycling, particularly given Dresden’s bike-friendly infrastructure. 

Furthermore, we conduct an analysis of all major motorway entrances and exits to identify traffic redistribution and new route preferences. Dresden’s motorways form a peripheral ring that supports both urban and regional connectivity, making them key indicators of systemic adaptation.

Table~\ref{tab:motorway_traffic_changes} presents the changes in traffic volumes across six key motorway access points. The most significant increase in inbound traffic is observed at Dresden Neustadt, which sees a surge of +3,497 vehicles, despite a moderate outbound decrease of –947, resulting in a net increase of +2,550 vehicles. This pattern implies that Dresden Neustadt becomes a major diversion route to enter the city post-collapse, likely due to its connectivity with alternative bridges or transit nodes. In contrast, Dresden Hellarau experiences the largest net decrease (–1,927 vehicles), with substantial drops in both inbound (–781) and outbound (–1,146) traffic. This indicates a possible avoidance of this node due to increased travel time or re-routing through more accessible alternatives. Similar patterns are observed at Dresden Altstadt (–1,062) and Südvorstadt (–1,029), both experiencing reductions in both directions, suggesting decreased network attractiveness in these areas. Meanwhile, Dresden Gorbitz also shows a moderate net loss (–624 vehicles), particularly in outbound traffic. Interestingly, Dresden Wilder Mann exhibits the smallest net change (–515 vehicles), implying a relatively stable yet slightly reduced role in commuter flow redistribution.

\begin{table}
\centering
\caption{Traffic volume changes at Dresden motorway nodes before and after the Carola Bridge collapse.}
\label{tab:motorway_traffic_changes}
\begin{tabular}{lrrr}
\toprule
           Location &  Inbound Change &  Outbound Change &  Net Change \\
\midrule
   Dresden Neustadt &            +3497 &             --947 &        +2550 \\
Dresden Wilder Mann &            --340 &             --175 &        --515 \\
   Dresden Hellarau &            --781 &            --1146 &       --1927 \\
   Dresden Altstadt &            --480 &             --582 &       --1062 \\
    Dresden Gorbitz &             --89 &             --535 &        --624 \\
Dresden Südvorstadt &            --761 &             --268 &       --1029 \\
\bottomrule
\end{tabular}
\end{table}

Overall, traffic across these key motorway nodes decreases by approximately 2,607 vehicles. This finding supports the hypothesis of modal shifts, alternative route choices, and decreased travel demand. Since motorways are typically used by commuters travelling from regions outside Dresden rather than inner-city residents, the observed decline suggests behavioural adjustment among long-distance travellers. While some bias may be introduced by outbound commuters from Dresden to other cities or by touristic traffic, their influence is likely limited, as such traffic is present in both the pre- and post-collapse datasets..

\section{Conclusion and Future Work}\label{sec:conclusion}
The sudden collapse of the Carola Bridge in Dresden triggered substantial disruptions across the city’s transport network, highlighting the vulnerability of urban infrastructure and the adaptive responses of travellers. Our study examined the spatio-temporal effects and traffic redistribution, changes in peak hours and behavioural adaptation following the incident, using high-resolution traffic data from key river crossings, arterial roads, P+R parking and motorway access points. Our analysis demonstrates that traffic redistribution is highly uneven, with major traffic shift toward the Albert Bridge, which experiences a 75\% surge in daily traffic volume, while the Marien Bridge, already operating at capacity, absorbs an additional 6,000 vehicles per day. The analysis of peak-hour dynamics reveals a substantial extension of peak period, with peak durations prolonged by up to 250 minutes. Overall, the network experiences a decline of around 8,000 vehicles per day, suggesting a broader behavioural adaptation that includes potential mode shifts, reduced trips, or adoption of alternative work patterns such as remote work. The observed increase in P+R usage supports the hypothesis that some travellers switch to public transport, while reduced activity at motorway access points, particularly in Hellarau and Südvorstadt, points to decreased commuter inflow from regional origins.

However, this study primarily focuses on traffic flow redistribution rather than the transport system as a whole, since the available detector data mainly capture motorised traffic count and speed. The lack of origin–destination data and explicit mode information limited our ability to quantify route choice adaptations or modal transitions in detail. Future research should extend this perspective to multimodal datasets, enabling the integration of public transport ridership, cycling, and pedestrian flows to provide a more complete picture of systemic adaptation. In addition, future research should be extended to investigate the longer-term impacts such as potential shifts in travel behaviour, land use patterns, and mode choice preferences. Finally, future work could explore resilient multimodal transport control strategies, for example by integrating macroscopic flow models such as the Cell Transmission Model (CTM) and Link Transmission Model (LTM) \citep{Wei2025}, to evaluate the performance of different control strategies under disruption scenarios and to support the design of more resilient urban transport systems.


\section*{Statement}
During the preparation of this manuscript, the authors used OpenAI ChatGPT in order to improve language in the writing process. After using this tool/service, the authors reviewed and edited the content as needed and took full responsibility for the content of the publication.

\bibliographystyle{elsarticle-harv}
\bibliography{references}

\appendix
\section{Mathematical Formulation of Statistical Tests}
This appendix provides the mathematical definitions and test statistics for the paired $T$-test and ANOVA test used in the methodological framework of this study.

\label{app:stat-analysis}
\subsection{Paired $T$-Test}
\label{app:t-test}
To assess whether the traffic flow exhibits a statistically significant change following the bridge collapse, a paired $T$-test is employed. In this study, the two paired samples represent traffic flow before (\(q_{\text{pre},i}\)) and after (\(q_{\text{post},i}\)) the bridge collapse, for each sensor \(i = 1, 2, \ldots, n\). The test statistic is given by:

\begin{equation}
t = \frac{\bar{d}}{s_d / \sqrt{n}}
\label{eq:t-test}
\end{equation}
\noindent
where $\bar{d} = \frac{1}{n}\sum_{i=1}^{n} d_i$ is the mean of the paired differences, $s_d = \sqrt{\frac{1}{n - 1}\sum_{i=1}^{n}(d_i - \bar{d})^2}$ is the standard deviation of the differences, $d_i = q_{\text{post},i} - q_{\text{pre},i}$ is the traffic flow difference for the $i^{\text{th}}$ sensor, and $n$ is the total number of paired observations. The statistic $t$ follows a $t$-distribution with $n - 1$ degrees of freedom. A significant change in traffic flow is inferred when the computed $|t|$ exceeds the critical value or when the associated $p$-value is less than 0.05.

\subsection{Analysis of Variance (ANOVA)}
\label{app:anova}
To examine whether mean traffic flow differs significantly across multiple groups (e.g., bridges or traffic corridors), an ANOVA test is applied. ANOVA tests the null hypothesis that all group means are equal against the alternative hypothesis that at least one group mean differs.The ANOVA test statistic is expressed as:

\begin{equation}
F = \frac{MS_B}{MS_W} = \frac{SS_B / (k - 1)}{SS_W / (N - k)}
\label{eq:anova}
\end{equation}
\noindent
where $SS_B = \sum_{j=1}^{k} n_j (\bar{q}_j - \bar{q})^2$ is the between-group sum of squares, 
$SS_W = \sum_{j=1}^{k}\sum_{i=1}^{n_j} (q_{ij} - \bar{q}_j)^2$ is the within-group sum of squares, $MS_B = SS_B / (k - 1)$ and $MS_W = SS_W / (N - k)$ are the corresponding mean squares, $k$ is the number of groups (e.g., bridges or corridors), $n_j$ is the number of observations in group $j$, and $N = \sum_{j=1}^{k} n_j$ is the total number of observations. The test statistic $F$ follows an $F$-distribution with $(k - 1, N - k)$ degrees of freedom. A significant difference among group means is inferred when the computed $F$ exceeds the critical value or when the $p$-value is less than 0.05.

\section{Lookup Table for Detector IDs and Figure Labels}
\label{app:look_up}

\begin{table}[H]
\centering
\renewcommand\arraystretch{0.96}
\captionsetup{font={small}}
\footnotesize
\begin{center}
  \caption{Information of short indices and figure labels to detector IDs for critical bridges, arteries, and corridors.}
  \vspace{1.5ex}
  \label{tab:critical-detectors-compact}
  \begin{tabular}{%
    >{\raggedright\arraybackslash}p{2.1cm}
    >{\raggedright\arraybackslash}p{3.6cm}
    >{\centering\arraybackslash}p{1.6cm}
    >{\centering\arraybackslash}p{1.8cm}
    >{\centering\arraybackslash}p{0.9cm}
    >{\centering\arraybackslash}p{2.5cm}
    >{\centering\arraybackslash}p{1.8cm}
  }
  \hline
  \textbf{Category} & \textbf{Name} & \textbf{Direction} & \textbf{Cardinal Direction} & \textbf{Index} & \textbf{Detector ID} & \textbf{Figure Label} \\
  \hline
  \multirow{5}{=}{Bridge} & \multirow{3}{=}{Marien bridge} & \multirow{2}{*}{outbound} & \multirow{2}{*}{SW\,--\,NE} & 1.1 & 1388 & MBU-1 \\
                          &                                  &                            &  & 1.2 & 1389 & MBU-2 \\
                          &                                  & inbound                    &  NE\,--\,SW & 2 & 1408\_1407 & MBI-1 \\
                          & \multirow{2}{=}{Albert bridge}   & outbound                   & S\,--\,N & 3 & 1431\_1432 & ABU-1 \\
                          &                                  & inbound                    & N\,--\,S & 4 & 1429\_1430 & ABI-1 \\
  \hline
  \multirow{21}{=}{Artery} & \multirow{7}{=}{Marien bridge-related artery} & \multirow{3}{*}{inbound}  & \multirow{3}{*}{N\,--\,S} & 5 & 1129\_1130 & MAI-1 \\
                           &                                             &                           &  & 6 & 1287 & MAI-2 \\
                           &                                             &                           &  & 7 & 911\_912 & MAI-3 \\
                           &                                             & \multirow{4}{*}{outbound} & \multirow{4}{*}{S\,--\,N} & 8 & 909\_910 & MAU-1 \\
                           &                                             &                           &  & 9 & 1394\_1395 & MAU-2 \\
                           &                                             &                           &  & 10 & 1416 & MAU-3 \\
                           &                                             &                           &  & 11 & 1128 & MAU-4 \\
                           \cline{2-7}
                           & \multirow{6}{=}{Carola bridge-related artery} & \multirow{3}{*}{inbound}  & \multirow{5}{*}{N\,--\,S} & 12 & 1310\_1311 & CAI-1 \\
                           &                                              &                           &  & 13 & 1401\_1402 & CAI-4 \\
                           &                                              &                           &  & 14 & 1319\_1320 & CAI-5 \\
                           &                                              & \multirow{3}{*}{outbound} & \multirow{4}{*}{S\,--\,N} & 15 & 1317\_1318 & CAU-1 \\
                           &                                              &                           &  & 16 & 1329\_1330 & CAU-2 \\
                           &&&&&\\
                           \cline{2-7}
                           & \multirow{5}{=}{Albert bridge-related artery} & \multirow{3}{*}{inbound}  & \multirow{3}{*}{N\,--\,S} & 17 & 1359 & AAI-1 \\
                           &                                              &                           &  & 18 & 1447\_1448 & AAI-2 \\
                           &                                              &                           &  & 19 & 1257 & AAI-3 \\
                           &                                              & \multirow{2}{*}{outbound} & \multirow{2}{*}{S\,--\,N} & 20 & 1274 & AAU-1 \\
                           &                                              &                           &  & 21 & 1379 & AAU-2 \\
  \hline
  \multirow{10}{=}{Corridor} & \multirow{2}{=}{Outer New City--Inner New City Corridor} & outbound & S\,--\,N & 22 & 1371 & OICU-1 \\
                             &                                                            & inbound  & N\,--\,S & 23 & 1322\_1323 & OICI-1 \\
                             \cline{2-7}
                             & \multirow{4}{=}{Leipzig Suburb--Inner New City Corridor}  & \multirow{1}{*}{outbound} & \multirow{1}{*}{S\,--\,N} & 24 & 1405\_1406 & LICU-1 \\
                             &                                                            & \multirow{3}{*}{inbound}   & \multirow{3}{*}{N\,--\,S} & 25 & 1308\_1309 & LICI-1 \\
                             &                                                            &                              &  & 26 & 1367\_1368 & LICI-2 \\
                             &                                                            &                              &  & 27 & 1396\_1397 & LICI-3 \\
                             \cline{2-7}
                             & \multirow{4}{=}{Inner New City East--West Corridor}       & \multirow{2}{*}{inbound}   & E\,--\,W & 28 & 1239 & IEI-1 \\
                             &                                                            &                              & W\,--\,E & 29 & 1361\_1360 & IEI-2 \\
                             &                                                            & \multirow{2}{*}{outbound}  & W\,--\,E & 30 & 1238 & IEU-1 \\
                             &                                                            &                              & E\,--\,W & 31 & 1392\_1393 & IEU-2 \\
  \hline
  \end{tabular}
\end{center}
\end{table}

\section{Analysis Peak-Hour}
\label{app1}

 \begin{figure}[H]
    \centering
    \begin{subfigure}[b]{0.48\textwidth}
        \centering
        \includegraphics[width=\textwidth]{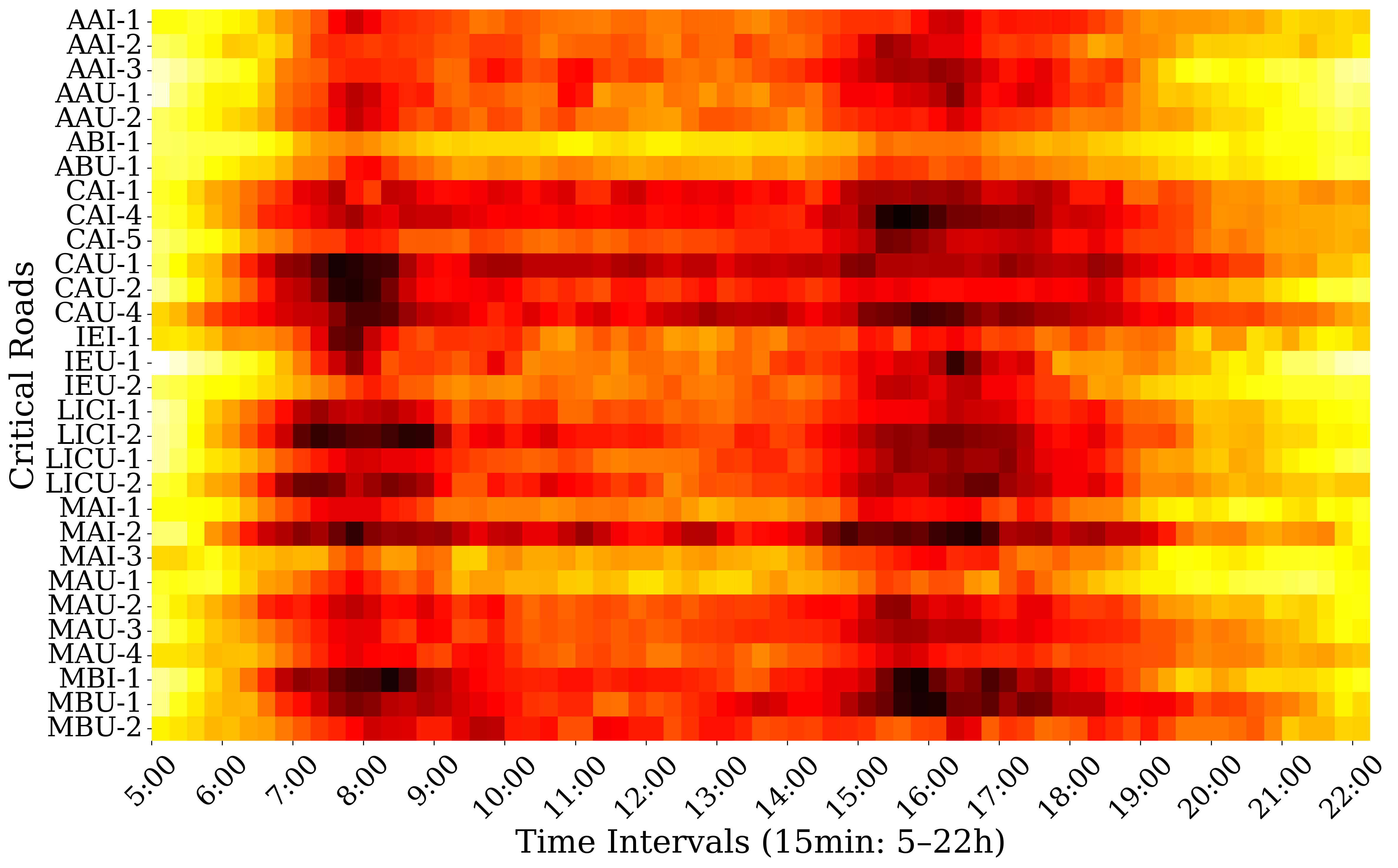}
        \caption{Tuesday before.}
        \label{fig:phi_before}
    \end{subfigure}
    \begin{subfigure}[b]{0.48\textwidth}
        \centering
        \includegraphics[width=\textwidth]{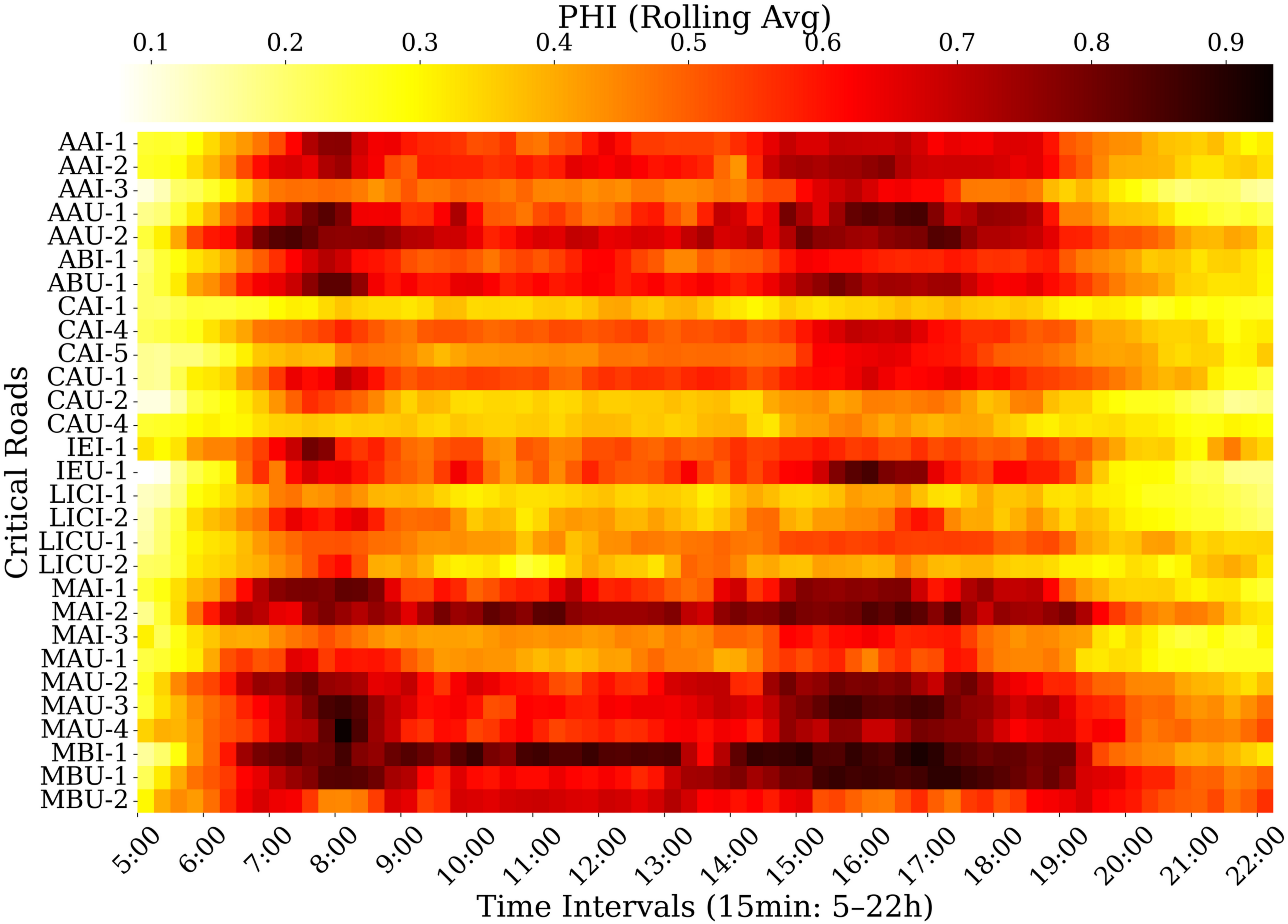}
        \caption{Tuesday after}
        \label{fig:phi_after}
    \end{subfigure}
    \begin{subfigure}[b]{0.48\textwidth}
        \centering
        \includegraphics[width=\textwidth]{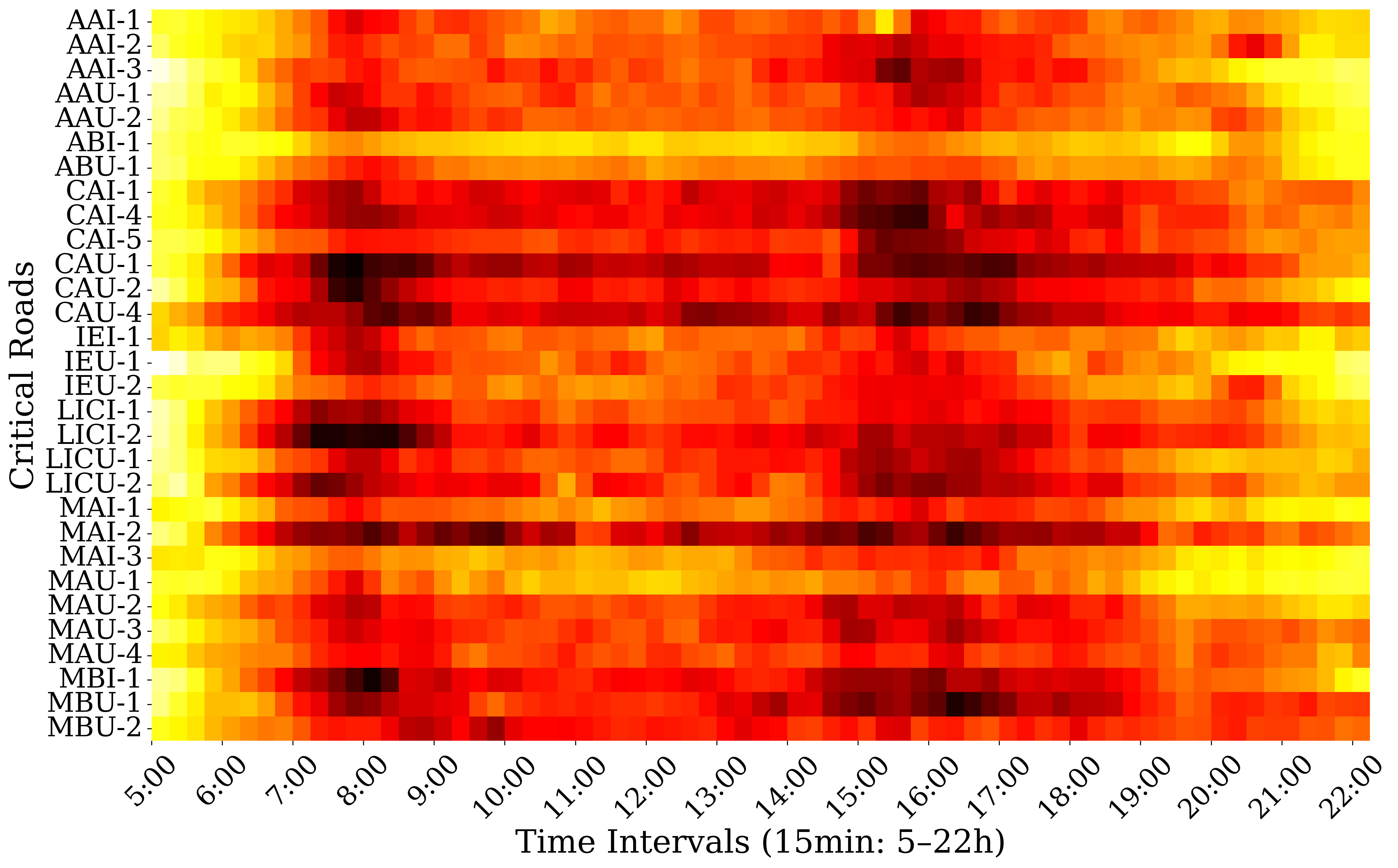}
        \caption{Thursday before.}
        \label{fig_phi_distribution}
    \end{subfigure}
    \begin{subfigure}[b]{0.48\textwidth}
        \centering
        \includegraphics[width=\textwidth]{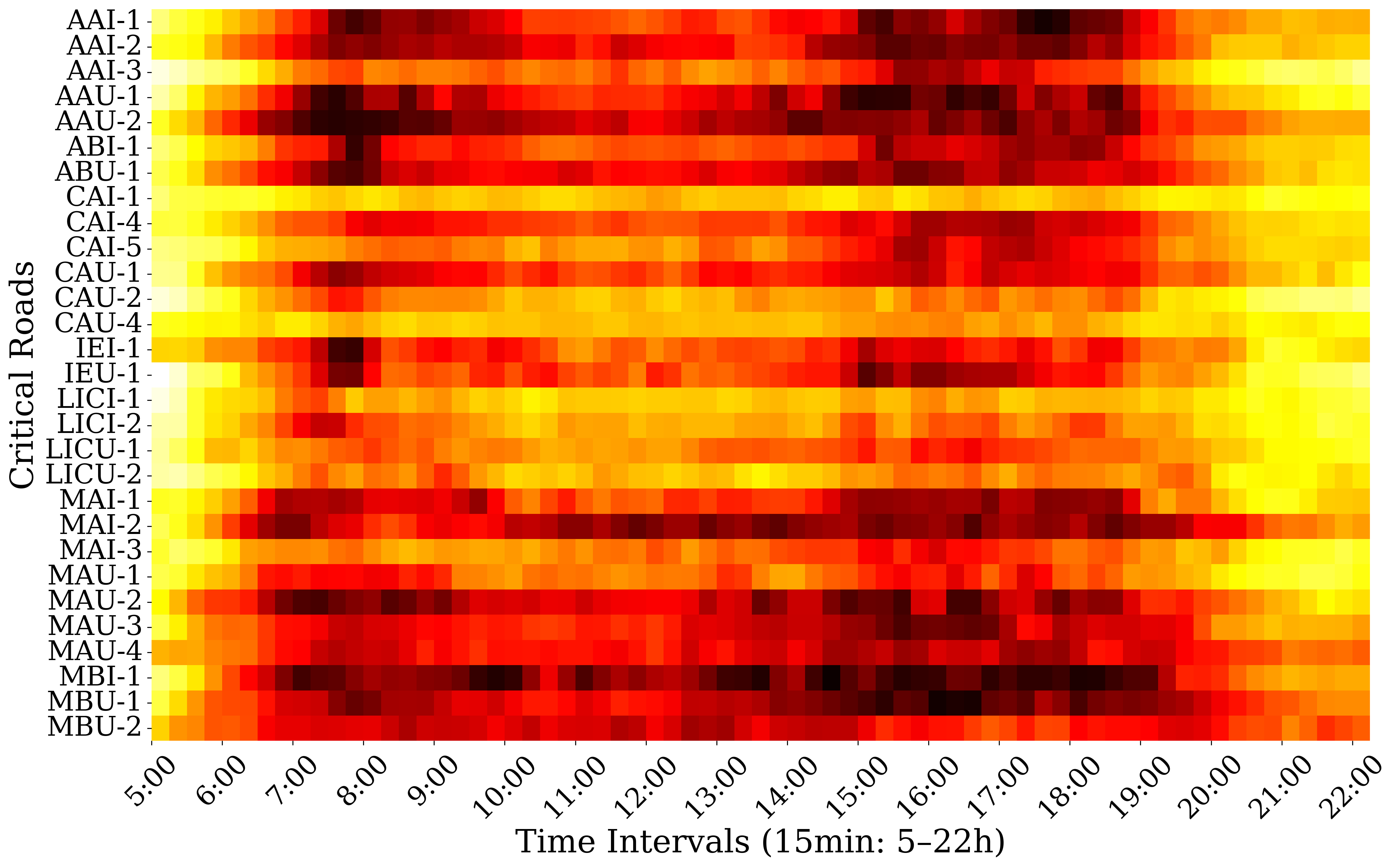}
        \caption{Thursday before}
        \label{fig:minutes_changes}
    \end{subfigure}
    
    \caption{Analysis of daily peak hour changes after the Carola bridge collapse. }
    \label{fig:phi_comparison_app}
\end{figure}

 \begin{figure}[H]
    \centering
    \begin{subfigure}[b]{0.7\textwidth}
        \centering
        \includegraphics[width=\textwidth]{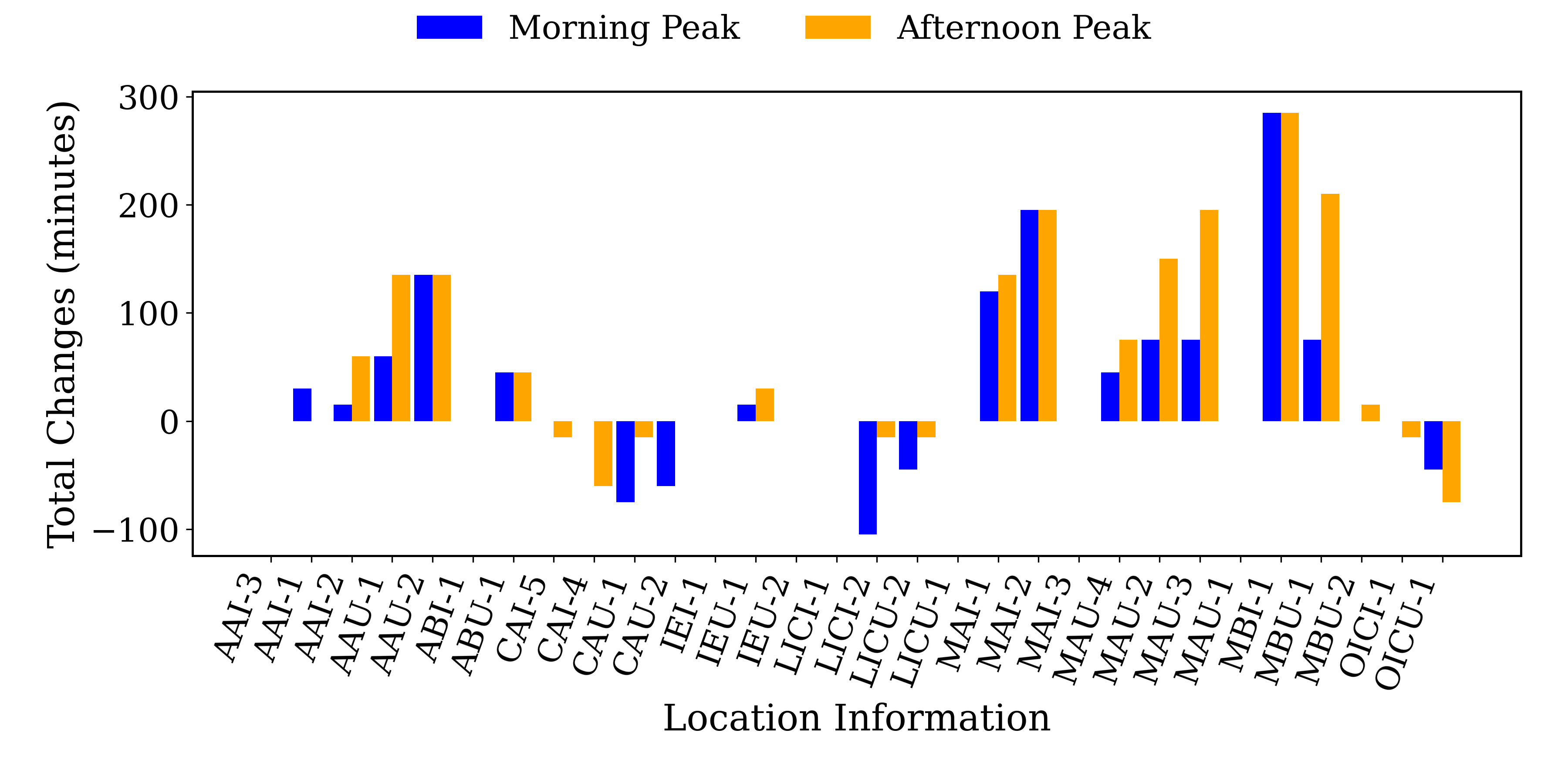}
        \caption{Tuesday Minute Changes.}
        \label{fig:total_peak_changes_phi_before}
    \end{subfigure}
    \begin{subfigure}[b]{0.7\textwidth}
        \centering
        \includegraphics[width=\textwidth]{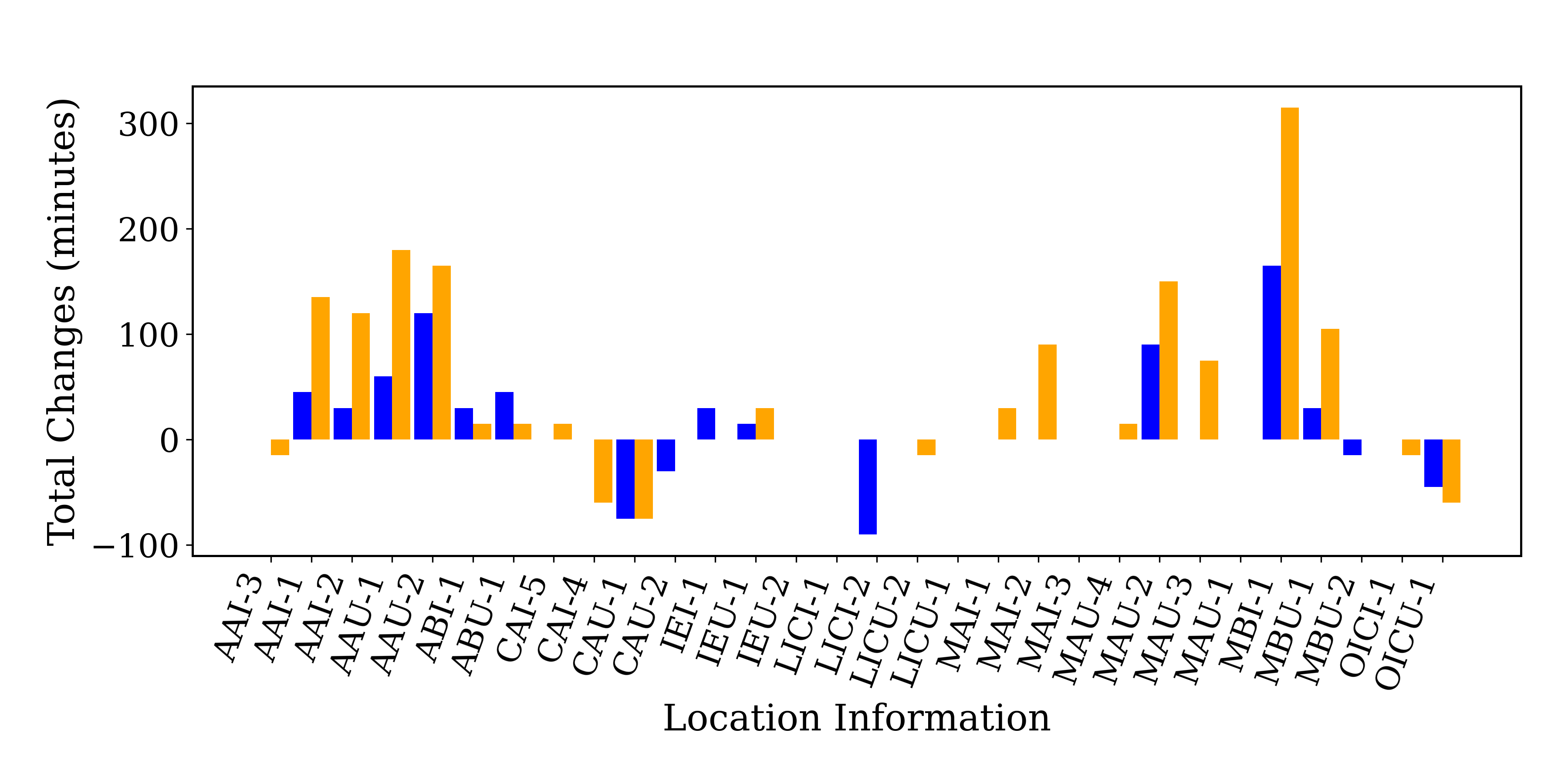}
        \caption{Thursday Minute Changes.}
        \label{fig:total_peak_changes_phi_after}
    \end{subfigure}
    \caption{Analysis of daily peak hour changes after the Carola bridge collapse. }
    \label{fig:minute_chnages_app}
\end{figure}

\section{Maximum Observed Operational Flow Calculation}
\label{app:moof}

\begin{figure}[H]
    \centering
    \includegraphics[width=0.98\linewidth]{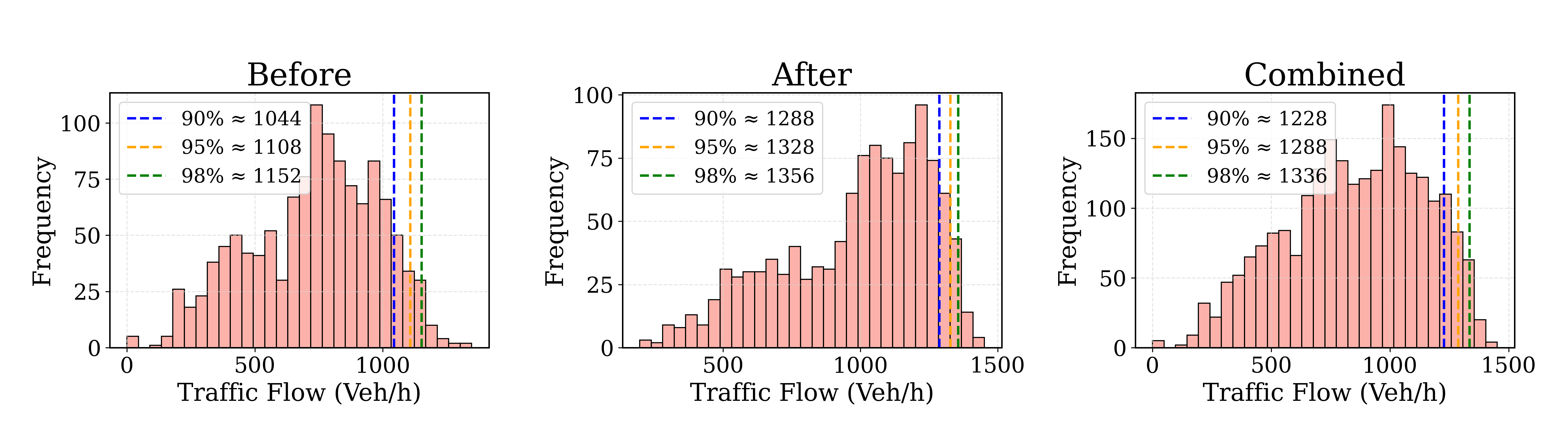}
    \caption{Distribution of hourly traffic flow (in Veh/h) across all 15-minute intervals before, after, and combined periods for the one-lane cross section. Vertical dashed lines indicate the 90th, 95th, and 98th percentile thresholds, respectively. The 95th percentile value from the combined period (1288 Veh/h) is adopted as the estimated maximum observed operational flow for this one-lane road segment.}
    \label{fig:capacity_one_lane}
\end{figure}

\begin{figure}[H]
    \centering
    \includegraphics[width=0.98\linewidth]{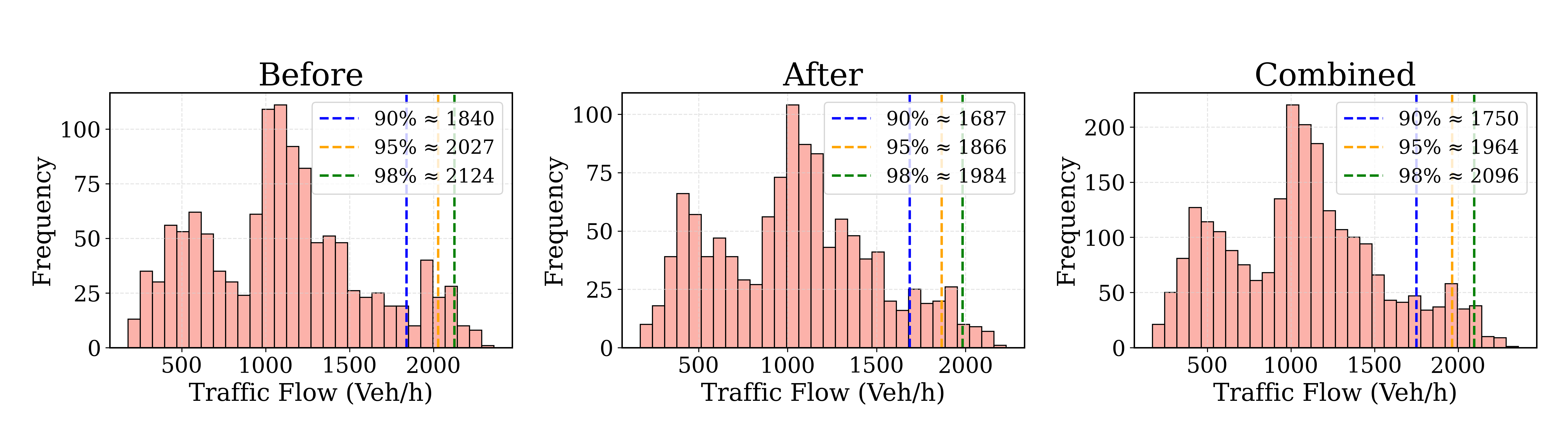}
    \caption{Distribution of hourly traffic flow (in Veh/h) across all 15-minute intervals before, after, and combined periods for the two-lane cross section. Vertical dashed lines indicate the 90th, 95th, and 98th percentile thresholds, respectively. The 95th percentile value from the combined period (1964 Veh/h) is adopted as the estimated maximum observed operational flow for this two-lane road segment.}
    \label{fig:capacity_two_lane}
\end{figure}

\end{document}